\documentclass[12pt]{article}
\usepackage{lscape}
\usepackage{setspace}
\usepackage{ifdraft}
\usepackage{multirow}
\usepackage[backend=biber, style=apa]{biblatex}

\usepackage{pdflscape}
\usepackage{color}
\usepackage{hyperref}
\hypersetup{colorlinks=false,pdfborderstyle={/S/U/W 1}}
            
\usepackage{amsmath, amssymb, url, bm}
\usepackage{rotating}
\usepackage{latexsym}
\usepackage{graphicx}
\usepackage{fancyvrb}  % Verbatim environment
\usepackage{caption}
\usepackage{subcaption}
\usepackage{indentfirst}
\newcommand\spacingset[1]{\renewcommand{\baselinestretch}{#1}\small\normalsize}

\newcommand*{\thisdraft}{July 8, 2023} % define command
\newcommand*{\firstdraft}{First draft: April 4, 2023}

\newtheorem{theorem}{Theorem}[section]

\newtheorem{lemma}[theorem]{Lemma}

\newtheorem{proposition}{Proposition}
\newtheorem{assumption}{Assumption}
\newtheorem{definition}{Definition}
\newcommand{\qed}{\nobreak \ifvmode \relax \else
      \ifdim\lastskip<1.5em \hskip-\lastskip
      \hskip1.5em plus0em minus0.5em \fi \nobreak
      \vrule height0.75em width0.5em depth0.25em\fi}

\DeclareMathOperator*\argmin{argmin}

\renewcommand{\epsilon}{\varepsilon}

\usepackage{tikz}
\usepackage{pgfpages}
\usetikzlibrary{shapes}
\usetikzlibrary{positioning}
\usetikzlibrary{arrows}
\usetikzlibrary{shapes.misc}
\usetikzlibrary{shapes.symbols}
\usetikzlibrary{shadows}
\usetikzlibrary{fit}

\usepackage{appendix}

\begin{document}

%%%%%%%%%%%%%%%%%%%%%%%%%%%%%%%%%%%%%%%%%%%%%%%%%%%%%%%%%%%%%%%%%%%%%%%%%%%%%%%%
% == Title Page
%%%%%%%%%%%%%%%%%%%%%%%%%%%%%%%%%%%%%%%%%%%%%%%%%%%%%%%%%%%%%%%%%%%%%%%%%%%%%%%%

\newcommand{\blind}{0}

    \newcommand{\tit}{Decomposing Triple-Differences Regression under Staggered Adoption}

\if0\blind
\title{\bf \tit}
  \author{Anton Strezhnev\thanks{Assistant Professor, University of Chicago Department of Political Science. Email: \href{mailto:astrezhnev@uchicago.edu}{astrezhnev@uchicago.edu}. I thank Andy Eggers, Justin Grimmer, Bobby Gulotty, Silvia Kim, Apoorva Lal, Molly Offer-Westort, Miguel Rueda, Yiqing Xu and Arthur Yu, as well as participants at the Stanford Political Science Methods Workshop, the NYU-AD Theory in Methods Workshop, and the 2023 American Causal Inference Conference for helpful discussions and comments.}}

\date{\thisdraft \\ \firstdraft}

\maketitle
\if\blind
\maketitle
\fi

\pdfbookmark[1]{Title Page}{Title Page}

\thispagestyle{empty}
\setcounter{page}{0}
\spacingset{1}

\begin{abstract}
The triple-differences (TD) design is a popular identification strategy for causal effects in settings where researchers do not believe the parallel trends assumption of conventional difference-in-differences (DiD) is satisfied. TD designs augment the conventional 2x2 DiD with a ``placebo" stratum -- observations that are nested in the same units and time periods but are known to be entirely unaffected by the treatment. However, many TD applications go beyond this simple 2x2x2 and use observations on many units in many ``placebo" strata across multiple time periods. A popular estimator for this setting is the triple-differences regression (TDR) fixed-effects estimator -- an extension of the common ``two-way fixed effects" estimator for DiD. This paper decomposes the TDR estimator into its component two-group/two-period/two-strata triple-differences and illustrates how interpreting this parameter causally in settings with arbitrary staggered adoption requires strong effect homogeneity assumptions as many placebo DiDs incorporate observations under treatment. The decomposition clarifies the implied identifying variation behind the triple-differences regression estimator and suggests researchers should be cautious when implementing these estimators in settings more complex than the 2x2x2 case. Alternative approaches that only incorporate ``clean placebos" such as direct imputation of the counterfactual may be more appropriate. The paper concludes by demonstrating the utility of this imputation estimator in an application of the ``gravity model" to the estimation of the effect of the WTO/GATT on international trade.
\end{abstract}

\clearpage

%%%%%%%%%%%%%%%%%%%%%%%%%%%%%%%%%%%%%%%%%%%%%%%%%%%%%%%%%%%%%%%%%%%%%%%%%%%%%%%%
% == Body
%%%%%%%%%%%%%%%%%%%%%%%%%%%%%%%%%%%%%%%%%%%%%%%%%%%%%%%%%%%%%%%%%%%%%%%%%%%%%%%%

\spacingset{1.5}

\section{Introduction}

Despite its growing popularity in applied work, the triple-differences design remains understudied. While differences-in-differences (DiD) designs are ubiquitous in applied causal research, their validity hinges on an assumption of ``parallel counterfactual trends" among treated and control groups. In the simple 2x2 setting with two time periods and a treated and control group, this assumption states that the treated group would have followed the same trend over time as the control group had the treated group instead received control. This may not hold if the types of units in the treated group are differentially exposed to some time-dependent shock. The triple-differences estimator attempts to address violations of this assumption by incorporating an additional difference-in-differences term that captures these shocks. In the simplest 2x2x2 setting, this can be understood as a difference between a ``primary" difference-in-differences and a ``placebo" difference-in-differences where the placebo DiD is constructed using observations that retain the same structure as the primary DiD but are known to be unaffected by the treatment. For example, \textcite{gruber1994incidence}, noted as one of the first studies to explicitly use a triple-differences strategy by \textcite{olden2020triple}, examines the effect of state-mandated maternity benefits on labor market outcomes. The primary analysis uses a conventional difference-in-differences design, examining outcomes among individuals at risk of having a child across different states over time. Some states expanded insurance coverage mandates for maternity care while others did not. This primary analysis is augmented by a second, placebo, difference-in-difference which leveraged the fact that other individuals who were unaffected by the treatment could be observed in the same states and same time periods -- individuals who are known to be incapable of becoming pregnant. Under the assumption that the violation of the parallel trends assumption in the ``primary" stratum (individuals considered at-risk of childbirth) is equivalent to the parallel trends violation in the ``placebo" stratum (individuals not at risk), subtracting the placebo difference-in-difference estimate from the primary difference-in-difference identifies an average treatment effect on the treated.

Triple-difference designs leverage a structural feature common to many datasets where units may belong to multiple overlapping sub-groups that differ in their exposure to treatment.\footnote{Another distinct use of a ``triple-differences" approach involves incorporating additional pre-treatment periods in a conventional differences-in-differences setting as in \textcite{egami2023using}. This paper does not address this estimator.} In the case of \textcite{gruber1994incidence} individuals are nested within both state and ``risk-for-pregnancy" groupings. In this case, the second grouping is binary and is known to \textit{never} receive the treatment. A similar approach is seen in \textcite{gingerich2019ballot} which examines the effect of the staggered introduction of the secret ballot on voter behavior in elections for Brazil's Chamber of Deputies. Here, the placebo stratum consists of elections to the Senate for which the ballot reforms had already been uniformly implemented. Likewise, \textcite{agan2018ban} study the effect of ``ban the box" policies -- policies that prevent employers from asking prospective employees about criminal records -- on racial discrimination. Here, the triple-difference ``placebo" group consists of those employers that never asked about criminal records even prior to the ban and are therefore plausibly unaffected by changes in the policy.

However, not all triple-differences designs restrict themselves to a single treated stratum and a single placebo stratum. A common triple-differences setting examines outcomes observed at the individual or firm level where the individual or firm is located in a particular state and also belongs to a particular industry. For example \textcite{marchingiglio2019employment} uses a triple-differences design to estimate the effect of gender-specific minimum wage laws in the early 20th century United States. These laws were implemented at the state level and typically targeted specific industries that employed a larger share of women. As a result, treatment adoption is jointly determined by both the state \textit{and} their industry of employment. Unlike the case of the binary placebo grouping in \textcite{gruber1994incidence} and \textcite{agan2018ban} where all but the first group are entirely unaffected by treatment, different industries are exposed to treatment at different times and across different states. In such a design, each industry contains its own, separate, staggered difference-in-difference design where the treatment assignment varies across state and time.\footnote{Equivalently, one can consider separate, state-specific difference-in-differences designs that leverage treatment variation across industry and time -- the designation of which dimension determines the different strata is arbitrary.}

In this more general staggered triple-differences setting, where the times at which treatment is initiated can vary across both of the overlapping sub-groups, researchers have typically relied on the ``triple-differences regression" (TDR) specification to estimate a single summary treatment effect parameter. This is a ``static" fixed effects regression with a unit-specific intercept and separate group-time fixed effects for each of the sub-groups. Notably, this regression specification appears in many settings that do not explicitly defend a triple-differences identification strategy. For example, the canonical gravity model regression (\cite{egger2003proper}; \cite{head2014gravity}) in research on international trade has precisely this structure with each unit $i$ consisting of a single dyad with a sender (or exporter) $s$ and receiver (or importer) $r$. Such models are frequently used to study the effects of particular dyadic-level interventions, such as trade agreements or border restrictions, on trade flows (e.g. \cite{gowa2013politics}; \cite{carter2020barriers}). In the trade setting, this regression models the outcome $Y_{it}$ for dyad $i$ at time $t$ using a three-way fixed effects specification:
\begin{align}\label{eqn:tdrIndiv}
    Y_{it} = \tau D_{it} + \alpha_{s(i) r(i)} + \gamma_{s(i),t} + \delta_{r(i), t} +  \epsilon_{it}
\end{align}
where $D_{it}$ is an indicator for whether dyad $i$ is under treatment at time $t$, $s(i)$ denotes the sender country associated with dyad $i$ and $r(i)$ denotes the receiver country associated with dyad $i$. 

Researchers interpret estimates of $\tau$ as an estimate of some average treatment effect. In the canonical 2x2x2 setting with a treated group and a control group, two time periods, and two overlapping strata, the coefficient on $\hat{\tau}$ is equivalent to the simple triple-differences estimator of the average treatment effect on the treated \parencite{olden2020triple}. However, it remains unclear whether this interpretation is valid under a more general data structure that allows for staggering in the adoption of treatment not only across time but also across each of the overlapping strata. Recent work on the two-way fixed effects (TWFE) estimator has shown that when treatment roll-out is staggered, the two-way fixed effects estimator can be biased for the average treatment effect on the treated even if parallel trends holds, unless additional strong effect homogeneity assumptions are made (\cite{de2020two}; \cite{goodman2021difference}).

This paper shows that similar problems arise when using the static triple-difference regression estimator under common staggered adoption designs. It develops a decomposition the style of \textcite{goodman2021difference} for this estimator and shows that the regression coefficient on the treatment can be partially decomposed into an average over 2x2x2 triple-difference terms. In each of these 2x2x2s, the \textit{first difference} consists of the difference between the outcome of a unit under treatment and another unit under control within a given time period and stratum. The \textit{second difference} is the difference in the outcome between those same units in the same stratum but in a time period where both are under the same treatment status (both control or both treated). These first two differences constitute the ``primary" 2x2 DiDs. As in the case of TWFE, similar problems of ``forbidden" comparisons arise due to the use of treated units in this second difference term (\cite{goodman2021difference}; \cite{borusyak2021revisiting}).

The \textit{third difference} in the 2x2x2 is between a ``primary" 2x2 DiD term and another 2x2 ``placebo" DiD involving the same units and time periods but in a different stratum with a different distribution of treatment. In the case of the two-stratum treated/placebo triple-difference \parencite[e.g.][]{gruber1994incidence}, this placebo will always be a 2x2 where all unit-time periods are under control as the only other stratum that could be matched to a primary DiD never receives treatment. These are valid placebos even under effect heterogeneity as they incorporate no treated units. However, when treatment can be staggered arbitrarily, the placebo can also consist of 2x2 comparisons in other strata where some or all of the units are under treatment. These placebos are invalid in that they identify a combination of the bias in the primary DiD due to the violation of parallel trends and differences in treatment effects across units, time periods and strata. Notably, some of the primary DiD terms can themselves also act as placebos for other DiDs. Moreover, the absence of treatment staggering within stratum does not suffice to eliminate all invalid terms as it does in the two-way fixed effects setting. Even if treatment is not staggered within any single stratum, if the staggering differs across strata such that the treatment initiation times vary by strata or if some units that are untreated in one stratum are treated in another\footnote{The one exception to this is the case of a ``pure placebo" stratum where no units are treated.} then the triple differences regression will remain be biased for a weighted average of treatment effects even when a stronger identifying assumption -- parallel trends -- holds.

This paper builds on the recent developments in the identification and estimation of average treatment effects under difference-in-differences design assumptions in settings where treatment adoption is staggered over time (e.g. \cite{de2020two}; \cite{goodman2021difference}; \cite{sun2021estimating}; \cite{borusyak2021revisiting}; \cite{callaway2021difference}; \cite{imai2021use}; \cite{dube2023local}). These closely related papers have all highlighted how commonly-used ``two-way fixed effects" estimators fail to identify average treatment effects when not all units that adopt treatment adopt it at the same time when parallel trends assumptions hold but constant effects assumptions do not. This paper is closest in motivation to \textcite{goodman2021difference} which focuses on the static ``two-way fixed effects" estimator and provides an explicit decomposition of this estimator in terms of the underlying 2x2 differences-in-differences that comprise it. This decomposition provides additional intuition for the source of bias in TWFE beyond the general problem of ``negative weights" and illustrates the conditions under which the bias due to invalid or ``forbidden" comparisons is likely to be large or small. Likewise, while existing work on differences-in-differences under staggered adoption has noted that ``negative weights" are also likely to be a problem in the triple-differences \parencite{borusyak2021revisiting}, there has not been an explicit characterization of the source of the bias nor a discussion of what factors will accentuate it. The decomposition in this paper provides an answer to both of these questions.

The remainder of the paper is organized as follows. Section \ref{sec:definition} sets out a general framework for defining target causal estimands within a staggered triple-differences design. It extends the ``group-time" ATT framework of \textcite{callaway2021difference} and clarifies the necessary assumptions under which a \textit{stratum-specific} group-time ATT can be identified non-parametrically by a 2x2x2 triple-difference. It provides an alternative interpretation of the triple-differences identifying assumption from \textcite{olden2020triple} that generalizes to the setting with many placebo strata and arbitrary treatment staggering. Specifically, the assumption allows for a violation of parallel trends between two treatment histories in one stratum, but requires that the violation is \textit{constant} across matching units and time periods in other strata. Section \ref{sec:decomposition} presents the central contribution of this paper, a decomposition of the conventional triple-differences regression estimator into its component 2x2x2 comparisons. It shows how many of these 2x2x2 triple-difference comparisons are valid only under a strong constant treatment effects assumption as observations under treatment are used to estimate some of the ``placebo" difference-in-difference terms. Finally, Section \ref{sec:application} compares the conventional triple differences regression to an alternative approach based on direct imputation of the counterfactual by replicating the analysis in \textcite{goldstein2007institutions} of the effect of the WTO/GATT on bilateral trade. It suggests that existing studies in international trade that rely on the popular ``gravity model" approach are likely biased for the average of group-time ATTs due to the presence of effect heterogeneity across time. Moreover, placebo tests using the imputation method suggest that the triple-differences identifying assumptions implied by the gravity model are likely violated in the case of WTO/GATT membership. The paper concludes with a discussion of how applied researchers should approach triple-differences regressions and what features of the data are likely to minimize or exacerbate the problems highlighted by the decomposition.

\section{Defining the triple-differences estimand} \label{sec:definition}

This section begins by defining the causal estimands of interest in the staggered triple-differences setting, following the framework of \textcite{callaway2021difference} for staggered differences-in-differences. Consider a set of $T$ time periods denoted by $t = 1, 2, \dotsc, T$. For each unit $i$ we observe both an indicator for whether that unit is under treatment at time $t$, denoted $D_{it}$, and an outcome denoted $Y_{it}$. Define a unit's \textit{treatment history} as the vector of all treatment assignments from time $t=1$ to $t=T$: $\vec{D_i} = \{D_{i1}, D_{i2}, \dotsc, D_{iT}\}$. Under staggered adoption, all units are under control at time $t=1$ and can enter into treatment at any point between $t = 2$ through $t=T$. I assume that units entering treatment do not exit treatment (Assumption 1 of \textcite{callaway2021difference}):

\begin{assumption}\label{a:noreversal} No treatment reversal

$D_{i1} = 0$ for all $i$. $D_{i{t-1}} = 1$ implies $D_{it} = 1$

\end{assumption}

In the triple-difference design, observations are nested within two groupings denoted $s \in \{1, 2, \dotsc, S\}$ and $r \in \{1, 2, \dotsc, R\}$. For example, firms $i$ can be nested within \textbf{s}tates $s$ and indust\textbf{r}ies $r$ or in studies of international trade, directed dyads can be defined by their \textbf{s}ender $s$ and \textbf{r}eceiver $r$ combination.  In the directed-dyad case, each unique combination of $s$ and $r$ identifies a single observation while in most other other settings, multiple units can be grouped under the same $s$ and $r$. Let $s(i)$ denote the function that returns the grouping $s$ to which unit $i$ belongs and $r(i)$ the function that returns the grouping $r$. In the standard triple-difference setting, I assume that all units with the same $s$ and $r$ have the same treatment history.

\begin{assumption}\label{a:novariationgivensr} No treatment variation conditional on $s$ and $r$

For any two units $i$, $j$, if $s(i) = s(j)$ and $r(i) = r(j)$, then $\vec{D_i} = \vec{D_j}$  

\end{assumption}

Following the approach in \textcite{callaway2021difference}, \textcite{sun2021estimating} and \textcite{borusyak2021revisiting} one can summarize each unit's ``treatment history" under the no-reversal assumption using a single scalar: the treatment initiation time. Let $G_i$ denote the time period when unit $i$ first initiates treatment. For units that remain under control for all time periods -- the ``never-treated" units - let $G_i = \infty$. 

Define potential outcomes $Y_{it}(g)$ as a function of the treatment initiation time.\footnote{The potential outcomes can be more generally defined in terms of the entire treatment history vector $\vec{d} = \{d_1, d_2, \dotsc, d_T\}$. However, in a staggered adoption design, the entire treatment history is summarized by the initiation time.} I make a standard SUTVA/Consistency assumption that the observed outcome $Y_{it}$ for a unit that initiates treatment at $G_i = g$ is equal to the potential outcome $Y_{it}(\vec{d})$.

\begin{assumption} Consistency/SUTVA \label{a:sutva}

$Y_{it} = Y_{it}(g)$ if $G_i = g$

\end{assumption}

While any unique combination of $s$ and $r$ might contain multiple observations indexed by $i$, when treatment is assigned only at the level of the joint grouping, it is also helpful to consider defining potential outcomes at the level of each unique $s$, $r$ and $t$ combination.\footnote{A similar question of aggregation and level of analysis arises when defining treatment effects for cluster-randomized experiments and selecting between analyzing either individual or the cluster-averaged analyses \parencite{su2021model}.} Let $Y_{srt} = \frac{1}{N_{srt}}\sum_{i: s(i) = s, r(i) = r} Y_{it}$ where $N_{sr}$ denotes the number of units with $s(i) = s$ and $r(i) = r$. In some settings, such as gravity models, each combination of $s$ and $r$ will uniquely identify a single observation while in others, multiple units may belong to a common $s$ and $r$. Because all of these units will have the same treatment history, it is still useful to consider them as a single observation. Under Assumption \ref{a:novariationgivensr} define the treatment initiation time for all units $i: s(i) = s, r(i) = r$ as $G_{sr}$. The joint-grouping potential outcome at time $t$ can be written as $Y_{srt}(g)$, which is connected to the observed outcomes by Assumptions \ref{a:novariationgivensr} and \ref{a:sutva}.

\begin{assumption} Consistency/SUTVA of aggregated outcomes \label{a:sutva-aggregate}

$Y_{srt} = Y_{srt}(g)$ if $G_{sr} = g$

\end{assumption}

For the remainder of this paper, I will work exclusively with the aggregated outcomes and treatment history: $Y_{srt}$ and $G_{sr}$ and will refer to the grouping defined by $s$ as the ``unit" of analysis and the grouping defined by $r$ as the ``stratum." However, this designation is purely for convenience and one can consider either the grouping denoted $s$ or the grouping denoted $r$ as the ``stratum" grouping.

In a typical two-period design, there is only one causal estimand: the average treatment effect on the treated in time period 2. However, under staggered adoption there are many possible effects that can be identified. \textcite{callaway2021difference} define the group-time ATT as the building block of all causal quantities of interest under staggered adoption.\footnote{\textcite{sun2021estimating} define the same quantity but call it the cohort average treatment effect on the treated or CATT.} This quantity corresponds to the average treatment effect at some time period $t$ among units that initiate treatment at time $g$. In a triple-differences setting with multiple strata, it is necessary to refine the group-time ATT further by conditioning on the stratum $r$. The \textit{conditional} group-time ATT is defined as:

\begin{definition} Conditional group-time ATT \label{d:grouptimeATT}
    \begin{align*}
        \text{ATT}_r(g,t) = \mathbb{E}[Y_{srt}(g) - Y_{srt}(\infty) | G_{sr} = g]
    \end{align*}

\end{definition}

This represents the average difference at time $t$ between the observed outcome among units in stratum $r$ that initiate treatment at time $g$ and the counterfactual outcome that would have been observed had those units instead never received treatment. From this building block, one can define aggregate quantities that summarize the conditional group-time ATT across different treatment initiation times $g$, observation times $t$ and strata $r$. The choice of how to aggregate depends on a researcher's ultimate quantity of interest \parencite{callaway2021difference}.

Identifying a given conditional group-time ATT requires additional assumptions on the potential outcomes. First, as in the conventional difference-in-differences setting, the presence of ``anticipation" effects is ruled out such that altering treatment in the future does not change the potential outcomes of a unit in the past.

\begin{assumption} No anticipation \label{a:noanticipation}

    For all $t < g$, $Y_{srt}(g) = Y_{srt}(\infty)$
    
\end{assumption}

This assumption states that the potential outcome observed at time $t$ for a unit that initiates treatment at $g$ would be the same as the potential outcome at time $t$ had that unit instead \textit{never} initiated treatment as long as $t$ is prior to $g$. Combined with Assumption \ref{a:sutva-aggregate} it implies that the observed outcome $Y_{srt}$ equals the potential outcome $Y_{srt}(\infty)$ for any unit-time combination under control $D_{srt} = 0$. Moreover, $ATT_{r}(g, t) = 0$ for any $t < g$. Note that is possible to weaken this assumption to allow for limited anticipation up to a known number of periods, following \textcite{callaway2021difference}.

I first consider identification under the conventional (conditional) parallel trends assumption conditioning on stratum $r$.

\begin{assumption} Conditional parallel trends \label{a:conditionalparalleltrends}

    For all $r$, $t \neq t^{\prime}$, $g \neq g^{\prime}$:
    \begin{align*}
        \mathbb{E}[Y_{srt}(\infty) - Y_{srt^{\prime}}(\infty) | G_{sr} = g] - \mathbb{E}[Y_{srt}(\infty) - Y_{srt^{\prime}}(\infty) | G_{sr} = g^{\prime}] = 0
    \end{align*}
    
\end{assumption}

This version of the assumption is closest to the general parallel trends assumption in \textcite{borusyak2021revisiting} and also discussed in \textcite{roth2022what}. It assumes that parallel trends hold across all time periods and across all treatment groups. Other definitions of the parallel trends assumption weaken this further by assuming parallel trends holds only with respect to the never-treated group $G_{sr} = \infty$ or restricting the time period to only a single pre-treatment period \parencite{callaway2021difference}.

In the multi-strata setting, conditional parallel trends also only needs to hold for those strata $r$ for which there exist any treated units as there are no group-time ATTs for strata that are pure placebos - that is, those strata $r$ which for all $s$, $G_{sr} = \infty$. Under Assumption \ref{a:conditionalparalleltrends}, identification of any conditional group-time ATT is straightforward. Each stratum that contains any treated units is effectively its own staggered difference-in-difference.

\begin{proposition} Identification under conditional parallel trends \label{prop:IDparallel}

Under assumptions \ref{a:noreversal}, \ref{a:novariationgivensr}, \ref{a:sutva-aggregate}, \ref{a:noanticipation} and \ref{a:conditionalparalleltrends}
\begin{align*}
    \text{ATT}_r(g,t) = \mathbb{E}[Y_{srt} - Y_{srt^*} | G_{sr} = g] -  \mathbb{E}[Y_{srt} - Y_{srt^*} | G_{sr} = g^{\prime}]
\end{align*}
    for any $g \le t$, $t^* < g$, $g^{\prime} > t$

\end{proposition}

The proof follows from the results of \textcite{callaway2021difference} conditioning on stratum $r$. However, when Assumption \ref{a:conditionalparalleltrends} does not hold, researchers may nevertheless be able to identify the treatment effect by making an alternative identifying assumption and using a triple-differences design. In the triple-differences setting, researchers instead assume that while conditional parallel trends may be violated, that violation is \textit{constant} across strata.

\begin{assumption} Constant violation of conditional parallel trends \label{a:constantviolation}

    For all $r \neq r^{\prime}$, $t \neq t^{\prime}$, $g \neq g^{\prime}$:
    \begin{align*}
        \mathbb{E}[Y_{srt}(\infty) - Y_{srt^{\prime}}(\infty) | G_{sr} = g] - \mathbb{E}[Y_{srt}(\infty) - Y_{srt^{\prime}}(\infty) | G_{sr} = g^{\prime}] = \\ \mathbb{E}[Y_{sr^{\prime}t}(\infty) - Y_{sr^{\prime}t^{\prime}}(\infty) | G_{sr} = g] - \mathbb{E}[Y_{sr^{\prime}t}(\infty) - Y_{sr^{\prime}t^{\prime}}(\infty) | G_{sr} = g^{\prime}]
    \end{align*}
    
\end{assumption}

This generalizes the identification assumptions from \textcite{olden2020triple} to settings with multiple strata where the treatment can be arbitrarily staggered in different strata. It is worth noting that Assumption \ref{a:constantviolation} is not weaker than Assumption \ref{a:conditionalparalleltrends} as the former places restrictions on the control potential outcomes in a stratum $r^{\prime}$ conditional on treatment assigned in stratum $r$ irrespective of the treatment distribution in $r^{\prime}$. Therefore it is possible for parallel trends to hold within any strata that receive treatment but for this assumption to fail due to a parallel trends violation in a ``pure placebo" stratum in which no unit receives treatment.

Under this assumption, one can identify the conditional group-time ATT in stratum $r$ from the observed data by appending a second difference-in-differences to the result from \ref{prop:IDparallel}

\begin{proposition} Identification under a constant violation of conditional parallel trends \label{prop:IDtripdiff}

Under assumptions \ref{a:noreversal}, \ref{a:novariationgivensr}, \ref{a:sutva-aggregate}, \ref{a:noanticipation} and \ref{a:constantviolation}
\begin{align*}
    \text{ATT}_r(g,t) = \bigg\{\mathbb{E}[Y_{srt} - Y_{srt^*} | G_{sr} = g] -  \mathbb{E}[Y_{srt} - Y_{srt^*} | G_{sr} = g^{\prime}]\bigg\} -\\
    \bigg\{\mathbb{E}[Y_{sr^{\prime}t} - Y_{sr^{\prime}t^*} | G_{sr} = g, G_{sr^{\prime}} > t ] -  \mathbb{E}[Y_{sr^{\prime}t} - Y_{sr^{\prime}t^*} | G_{sr} = g^{\prime},  G_{sr^{\prime}} > t]\bigg\}
\end{align*}
    for any $g \le t$, $t^* < g$, $g^{\prime} > t$
    
\end{proposition}

The second difference-in-difference is comprised of those observations in stratum $r^{\prime}$ on the same units $s$ that are under control ($G_{sr^{\prime}} > t$) at both time $t$ and time $t^*$. Notably, the identification result conditions on \textit{both} the observed treatment in $r$ and $r^{\prime}$ in selecting valid units for the ``placebo" difference-in-differences. Plugging in sample analogues yields a consistent estimator for each conditional group-time ATT. Crucially, all of the observations used as part of the second difference in the primary DiD and the placebo DiD are under control even if they may initiate treatment in the future.

In practice, there may be very few observations available to estimate any individual $\text{ATT}_r(g,t)$  as is the case in conventional staggered DiD. Moreover, researchers are unlikely to be interested in a single, specific $\text{ATT}_r(g,t)$ but will instead target an ``average" effect for the sample. Following the approach of \textcite{callaway2021difference}, the conditional group-time ATTs can be aggregated using researcher-specified weights into a single treatment effect summary. For example, one could consider estimands that take the form of a weighted average across the non-zero stratum-specific group-time ATTs
\begin{equation}\label{eqn:aggregate}
    \text{ATT} = \sum_{r=1}^R \sum_{g=2}^T \sum_{t=g}^T \text{ATT}_{r}(g,t) w_{rgt}
\end{equation}

where $w_{rgt}$ denotes the weight assigned to the group-time ATT for group $g$ at time $t$ in stratum $r$. The choice of weights reflects researcher preferences over how to aggregate heterogeneous effects across different types of units (e.g. late versus early adopters) and over time (e.g. instantaneous versus longer-term effects). 

\section{Decomposing triple-differences regression} \label{sec:decomposition}

With a clearer understanding of the triple-differences estimand, it is worth asking whether the commonly-used triple-differences regression estimator identifies \textit{any} weighted average of stratum-specific group-time ATTs and what additional assumptions are required. Recent results decomposing the two-way fixed effects estimator \parencite{goodman2021difference} note that treatment effect heterogeneity results in bias due to invalid difference-in-differences comparisons. Intuitively, the TWFE estimator incorporates both valid 2x2 differences-in-differences - using \textit{past} time periods when treated and control units were under control to de-bias the cross-sectional first difference - as well as \textit{invalid} differences-in-differences - using \textit{future} time periods when treated and control units are both exposed to treatment. These ``invalid" terms appear when treatment adoption is staggered and late adopters act as controls for early adopters in earlier time periods. This section examines whether similar problems arise with the triple-differences regression by developing a decomposition of the estimator into its component 2x2x2 triple-difference comparisons. It shows that bias arises due to both invalid primary DiDs -- as in the two-way fixed effects case -- as well as invalid ``placebo" DiDs -- a feature unique to the triple-differences regression. 

As noted in \textcite{olden2020triple}, the most widely used regression triple-differences specification (equation \ref{eqn:tdrIndiv}) appears to come from a discussion of \textcite{yelowitz1995medicaid} in \textcite{angrist2008mostly} - a study of Medicaid expansion across states. In this section, I decompose a slightly different specification in which the outcome is aggregated among observations in the same $s$ and $r$ groupings. The specification regresses the aggregated outcome on an indicator for whether that unit-stratum is under treatment at time $t$ ($D_{srt}$) and three sets of fixed effects parameters: $\alpha_{sr}$ - the joint unit-stratum fixed effects,  and $\gamma_{st}$ - the unit-time fixed effects, and $\delta_{rt}$ - the stratum-time fixed effects. I assume a balanced panel with no missing observations.
\begin{align}\label{eqn:groupRTD}
    Y_{srt} = \tau D_{srt} + \alpha_{sr} + \gamma_{st} + \delta{rt} + \epsilon_{srt}
\end{align}

As discussed in the previous section, when there is more than one observation in a single $s,r,t$ cell, it can be shown that a weighted version of this regression where the weights on each observation are proportional to $N_{sr}$ yields an estimated $\hat{\tau}$ equivalent to that of Equation \ref{eqn:tdrIndiv}). In other words, the primary differences between \ref{eqn:tdrIndiv} and \ref{eqn:groupRTD} are in the different weights placed on each group-time treatment effect.

The decomposition relies on the application of the Frisch-Waugh-Lovell theorem to obtain an expression for the OLS estimator of $\tau$, $\hat{\tau}$ in terms of different averages of $Y$ across the groupings of the data: $s$, $r$, and $t$ (\cite{frisch1933partial}; \cite{lovell1963seasonal}). First, define the one-way, two-way and grand means of $Y$:

\begin{align}
    \bar{Y}_{sr} &\equiv \frac{1}{T} \sum_{t^{\prime}=1}^T Y_{srt^{\prime}} &
    \bar{Y}_{st} &\equiv  \frac{1}{R} \sum_{r^{\prime}=1}^R Y_{sr^{\prime}t} &
    \bar{Y}_{rt} &\equiv  \frac{1}{S} \sum_{s^{\prime}=1}^S Y_{s^{\prime}rt}\\
    \bar{\bar{Y}}_{s} &\equiv  \frac{1}{RT} \sum_{t^{\prime}=1}^T \sum_{r^{\prime}=1}^R Y_{sr^{\prime}t^{\prime}} &
    \bar{\bar{Y}}_{r} &\equiv  \frac{1}{ST} \sum_{t^{\prime}=1}^T \sum_{s^{\prime}=1}^S Y_{s^{\prime}rt^{\prime}} & 
    \bar{\bar{Y}}_{t} &\equiv  \frac{1}{SR} \sum_{s^{\prime}=1}^S \sum_{r^{\prime}=1}^R Y_{s^{\prime}rt^{\prime}}\\
    \bar{\bar{\bar{Y}}} &\equiv  \frac{1}{SRT} \sum_{s^{\prime}=1}^S \sum_{r^{\prime}=1}^R \sum_{t^{\prime}=1}^T Y_{s^{\prime}r^{\prime}t^{\prime}}
\end{align}

Define the grand means over the treatment indicator $D_{srt}$ analogously. The decomposition starts by applying the Frisch-Waugh-Lovell theorem and re-arranging the sums to write the OLS estimator $\hat{\tau}$ in terms of an average of the observed outcome $Y_{srt}$ in all treated unit-stratum-times $D_{srt} = 1$ that has been triple de-meaned.

\begin{lemma}\label{lem:ols}

The OLS estimator $\hat{\tau}$ in the grouped triple-differences regression can be written as:

\begin{align*}
   \hat{\tau} &= \frac{\sum_{s=1}^S \sum_{r=1}^R \sum_{t:D_{srt} = 1} Y_{srt} - \bar{Y}_{sr} - \bar{Y}_{st} - \bar{Y}_{rt} + \bar{\bar{Y}}_s + \bar{\bar{Y}}_r + \bar{\bar{Y}}_t - \bar{\bar{\bar{Y}}}}{ \sum_{s=1}^S \sum_{r=1}^R \sum_{t=1}^T \bigg(D_{srt} - \bar{D}_{sr} - \bar{D}_{st} - \bar{D}_{rt} + \bar{\bar{D}}_s + \bar{\bar{D}}_r + \bar{\bar{D}}_t - \bar{\bar{\bar{D}}}\bigg)^2}
\end{align*}

\end{lemma}

The terms in the numerator can be further rearranged by expanding out the double sums and obtaining an expression for $\hat{\tau}$ in terms of a weighted average of 2x2x2 comparisons.
\begin{align*}
   \hat{\tau} &= \frac{\sum\limits_{r=1}^R \sum\limits_{s=1}^S \sum\limits_{t=1}^T \sum\limits_{s^\prime=1}^S \sum\limits_{t^\prime=1}^T \sum\limits_{r^{\prime} \neq r} D_{srt}\bigg[\left(Y_{srt} - Y_{srt^{\prime}} - Y_{s^{\prime}rt} + Y_{s^{\prime}rt^{\prime}}\right) - \left(Y_{sr^{\prime}t} - Y_{sr^{\prime}t^{\prime}} - Y_{s^{\prime}r^{\prime}t} + Y_{s^{\prime}r^{\prime}t^{\prime}}\right)\bigg]}{ SRT \sum\limits_{s=1}^S \sum\limits_{r=1}^R \sum\limits_{t=1}^T \bigg(D_{srt} - \bar{D}_{sr} - \bar{D}_{st} - \bar{D}_{rt} + \bar{\bar{D}}_s + \bar{\bar{D}}_r + \bar{\bar{D}}_t - \bar{\bar{\bar{D}}}\bigg)^2}
\end{align*}

In this expression, $Y_{srt} - Y_{srt^{\prime}} - Y_{s^{\prime}rt} + Y_{s^{\prime}rt^{\prime}}$ is a 2x2 difference-in-difference within stratum $r$ (between unit $s$ and $s^\prime$ and time periods $t$ and $t^\prime$) while $Y_{sr^{\prime}t} - Y_{sr^{\prime}t^{\prime}} - Y_{s^{\prime}r^{\prime}t} + Y_{s^{\prime}r^{\prime}t^{\prime}}$ is the same difference-in-difference comparison in stratum $r^\prime$. Intuitively, the triple-differences regression yields an average over every such triple-difference where the first observation ($Y_{srt}$) is treated. However, the remaining seven observations are not guaranteed to be controls, resulting in invalid triple-difference terms which drive the bias due to heterogeneous effects. Enumerating every possible combination of treatment and control for the remaining seven terms yields the full triple-differences decomposition in Theorem \ref{thm:decomp}.

\begin{theorem} Regression triple-differences decomposition \label{thm:decomp}
    
    Let $D_{srt}^{(0)} \equiv (1 - D_{srt})$

    Define the difference-in-difference in stratum $r$, $\tilde{Y}_{srt}^{(s^{\prime}t^{\prime})}$ as
   \begin{align*}
       \tilde{Y}_{srt}^{(s^{\prime}r^{\prime}t^{\prime})} &\equiv Y_{srt} - Y_{s^{\prime}rt} - Y_{srt^{\prime}} + Y_{s^{\prime}rt^{\prime}}
    \end{align*}
    
    The regression triple-difference estimator can be written as:
   \begin{align*} 
   \hat{\tau} &= \omega^{-1} \sum_{r=1}^R \sum_{s=1}^S \sum_{t=1}^T \sum_{s^{\prime}=1}^S \sum_{t^{\prime}=1}^T \sum_{r^{\prime} \neq r} \bigg[\tilde{Y}_{srt}^{(s^{\prime}t^{\prime})} - \tilde{Y}_{sr^\prime t}^{(s^{\prime}t^{\prime})}  \bigg] \times \bigg[\underbrace{D_{srt}D_{s^{\prime}rt}^{(0)}D_{srt^{\prime}}^{(0)}D_{s^{\prime}rt^{\prime}}^{(0)}}_{\text{valid primary DiD}} + \underbrace{D_{srt}D_{s^{\prime}rt}^{(0)}D_{srt^{\prime}}D_{s^{\prime}rt^{\prime}}}_{\text{{invalid} primary DiD}}\bigg] \times \\ & \bigg[\underbrace{D_{sr^{\prime}t}^{(0)}D_{s^{\prime}r^{\prime}t}^{(0)}D_{sr^{\prime}t^{\prime}}^{(0)}D_{s^{\prime}r^{\prime}t^{\prime}}^{(0)}}_{\text{valid ``placebo" DiD}} + \underbrace{D_{sr^{\prime}t}^{(0)}D_{s^{\prime}r^{\prime}t}^{(0)}D_{sr^{\prime}t^{\prime}}D_{s^{\prime}r^{\prime}t^{\prime}}}_{\text{{invalid} ``placebo" DiD}} + \underbrace{D_{sr^{\prime}t}^{(0)}D_{s^{\prime}r^{\prime}t}D_{sr^{\prime}t^{\prime}}^{(0)}D_{s^{\prime}r^{\prime}t^{\prime}}}_{\text{{invalid} ``placebo" DiD}} + \\ &
   \underbrace{D_{sr^{\prime}t}D_{s^{\prime}r^{\prime}t}D_{sr^{\prime}t^{\prime}}D_{s^{\prime}r^{\prime}t^{\prime}}}_{\text{{invalid} ``placebo" DiD}} + \underbrace{D_{sr^{\prime}t}D_{s^{\prime}r^{\prime}t}D_{sr^{\prime}t^{\prime}}^{(0)}D_{s^{\prime}r^{\prime}t^{\prime}}^{(0)}}_{\text{{invalid} ``placebo" DiD}} + \underbrace{D_{sr^{\prime}t}D_{s^{\prime}r^{\prime}t}^{(0)}D_{sr^{\prime}t^{\prime}}D_{s^{\prime}r^{\prime}t^{\prime}}^{(0)}}_{\text{{invalid} ``placebo" DiD}} \bigg] +\\
   &  \bigg[ \tilde{Y}_{srt}^{(s^{\prime}t^{\prime})} - \tilde{Y}_{sr^\prime t}^{(s^{\prime}t^{\prime})}\bigg]\times \bigg[\underbrace{D_{srt}D_{s^{\prime}rt}^{(0)}D_{srt^{\prime}}^{(0)}D_{s^{\prime}rt^{\prime}}^{(0)}}_{\text{valid primary DiD}}\bigg] \times \bigg[ \underbrace{D_{sr^{\prime}t}^{(0)}D_{s^{\prime}r^{\prime}t}D_{sr^{\prime}t^{\prime}}^{(0)}D_{s^{\prime}r^{\prime}t^{\prime}}^{(0)} + D_{sr^{\prime}t}D_{s^{\prime}r^{\prime}t}D_{sr^{\prime}t^{\prime}}D_{s^{\prime}r^{\prime}t^{\prime}}^{(0)}}_{\text{matching ``flipped" DiDs in } r^{\prime}}\bigg] +\\
   &  \bigg[ \tilde{Y}_{srt}^{(s^{\prime}t^{\prime})} - \tilde{Y}_{sr^\prime t}^{(s^{\prime}t^{\prime})} \bigg]\times \bigg[\underbrace{D_{srt}D_{s^{\prime}rt}^{(0)}D_{srt^{\prime}}D_{s^{\prime}rt^{\prime}}}_{\text{{invalid} primary DiD}}\bigg] \times \bigg[ \underbrace{D_{sr^{\prime}t}^{(0)}D_{s^{\prime}r^{\prime}t}D_{sr^{\prime}t^{\prime}}D_{s^{\prime}r^{\prime}t^{\prime}} + D_{sr^{\prime}t}^{(0)}D_{s^{\prime}r^{\prime}t}^{(0)}D_{sr^{\prime}t^{\prime}}D_{s^{\prime}r^{\prime}t^{\prime}}^{(0)}}_{\text{matching ``flipped" DiDs in } r^{\prime}}\bigg]
   \end{align*}
   
   where the normalizing constant $\omega$ is equivalent to 
   \begin{align*}
       \omega \equiv SRT(N^{(1)}) &- SR\sum_{s=1}^S\sum_{r=1}^R (N_{sr}^{(1)})^2  -  ST\sum_{s=1}^S\sum_{t=1}^T (N_{st}^{(1)})^2 - RT\sum_{r=1}^R\sum_{t=1}^T (N_{rt}^{(1)})^2\\
    & + T \sum_{t=1}^T (N_{t}^{(1)})^2  + R \sum_{r=1}^R (N_{r}^{(1)})^2  + S \sum_{s=1}^S (N_{s}^{(1)})^2\\
    & - (N^{(1)})^2
   \end{align*}
   
  with $N^{(1)}$ and subscripts denoting the number of treated units in a particular set of strata.
\begin{align*}
    & N^{(1)}_{sr} = \sum_{t^{\prime}=1}^T D_{srt^{\prime}} && N^{(1)}_{st} = \sum_{r^{\prime}=1}^R D_{sr^{\prime}t} && N^{(1)}_{rt} = \sum_{s^{\prime}=1}^S D_{s^{\prime}rt}\\
    & N^{(1)}_{s} = \sum_{t^{\prime}=1}^T \sum_{r^{\prime} = 1}^R D_{sr^{\prime}t^{\prime}} && N^{(1)}_{r} = \sum_{t^{\prime}=1}^T \sum_{s^{\prime} = 1}^S D_{s^{\prime}rt^{\prime}} && N^{(1)}_{t} =  \sum_{s^{\prime} = 1}^S \sum_{r^{\prime}=1}^R D_{s^{\prime}r^{\prime}t}\\
    & N^{(1)} = \sum_{s^{\prime}=1}^S \sum_{r^{\prime} = 1}^R \sum_{t^{\prime}=1}^T  D_{s^{\prime}r^{\prime}t^{\prime}} 
\end{align*}

\end{theorem}

Theorem \ref{thm:decomp} makes clear that the regression triple differences estimator can be understood as a weighted average over two types of 2x2x2 terms: conventional triple-differences and triple-differences with invalid placebos. For each of these constituent terms, the treatment status of each unit-time-stratum is captured in eight treatment indicators: $D_{srt}$, $D_{s^\prime rt}$, $D_{srt^\prime}$ and $D_{s^\prime r t^\prime}$ for the first DiD in stratum $r$ and $D_{s r^\prime t}$, $D_{s^\prime r^\prime t}$, $D_{sr^\prime t^\prime}$ and $D_{s^\prime r^\prime t^\prime}$ for the second placebo DiD in stratum $r^\prime$. Each of these defines a particular triple-differences comparison, but only one of the possible comparisons comprises a valid triple-difference term that identifies a particular group-time ATT.

The first thing to note from the decomposition is that the same problems due to ``forbidden" comparisons in the two-way fixed effects estimator also appear here (\cite{de2020two}; \cite{goodman2021difference}; \cite{borusyak2021revisiting}). For each triple-difference, there are two types of primary DiD terms -- one involves a treated observation $D_{srt} = 1$ and three control observations ($D_{s^\prime r t} = 0, D_{s r t^\prime} = 0, D_{s^\prime r t^\prime}$ = 0) while the other involves a first difference between a treated observation $D_{srt} = 1$ and a cross-sectional control $D_{s^\prime rt} = 0$ subtracted from a second difference in time period $t^\prime$ where \textit{both} units $s$ and $s^\prime$ are under treatment ($D_{s r t^\prime = 1}$, $D_{s^\prime r t^\prime} = 1$). The first of these 2x2 comparisons identifies an ATT under parallel trends (Assumption \ref{a:conditionalparalleltrends}) or identifies an ATT plus a bias term under the ``constant violation of parallel trends" assumption (Assumption \ref{a:constantviolation}). The second 2x2 does \textit{not} identify an ATT even under parallel trends without an additional constant effects assumption. This is because under staggered adoption, the second difference involves two observations in a time period $t^\prime$ that follows $t$ ($t^\prime \ge G_{sr}$, $t^\prime \ge G_{s^\prime r}$)) where both $sr$ and $s^\prime r$ have adopted treatment. While the no anticipation assumption (Assumption \ref{a:noanticipation}) ensures that $\text{ATT}_r(G_{sr}, t^\prime) = \text{ATT}_r(G_{s^\prime r}, t^\prime) = 0$ for $t^\prime < G_{sr}$, $t^\prime < G_{s^\prime r}$, it does not guarantee this for $t^\prime \ge G_{sr}$. As a consequence, the difference in observed outcomes at $t^\prime$ between the two treated units incorporates a difference in two treatment effects, making it an invalid second difference term. All of the complications highlighted by \textcite{goodman2021difference} for TWFE are clearly still present in the triple-differences regression decomposition as well.

However, in the triple-differences setting, the problem of invalid comparisons is considerably more acute as even when all primary difference-in-difference terms are valid, the placebo difference-in-differences may not be. Among the set of placebo DiD terms that can appear in a given triple-difference, only one identifies the bias in the primary DiD under the constant violation of parallel trends assumption alone. This second DiD involves exclusively \textit{control} observations in stratum $r^\prime$: $D_{sr^\prime t} = D_{s^\prime r^\prime t} = D_{s r^\prime t^\prime} = D_{s^\prime r^\prime t^\prime} = 0$. All other placebo terms incorporate some observations that are treated. Five of these invalid second-differences incorporate an even number of treated observations such that the treatment effect, if constant, cancels and only a bias term remains. One such invalid second-DiD uses four treated observations instead of four control observations: $D_{sr^\prime t} = D_{s^\prime r^\prime t} = D_{s r^\prime t^\prime} = D_{s^\prime r^\prime t^\prime} = 1$. Another consists of either two control units at time $t$ ($D_{sr^\prime t} = D_{s^\prime r^\prime t} = 0$) and two treated units at $t^\prime$ ($D_{sr^\prime t^\prime} = D_{s^\prime r^\prime t^\prime} = 1$) or two treated units at time $t$ ($D_{sr^\prime t} = D_{s^\prime r^\prime t} = 1$) and two control units at time $t^\prime$ ($D_{sr^\prime t^\prime} = D_{s^\prime r^\prime t^\prime} = 0$). The last involves one unit $s$ that is treated at both $t$ and $t^\prime$ ($D_{s r^\prime t} = D_{s r^\prime t^\prime} = 1$) and another unit $s^\prime$ that is under control at both times  ($D_{s^\prime r^\prime t} = D_{s^\prime r^\prime t^\prime} = 0$) or vice-versa.

Intuitively, all of these placebo DiDs fail to identify the common violation of parallel trends under unrestricted effect heterogeneity. This is again because units $s r^\prime$ and unit $s^\prime r^\prime$ likely differ in their treatment initiation time $G_{s^\prime r} \neq G_{s^\prime r^\prime}$ and as a result, these placebo DiDs each involve differences in treatment effects across different timing groups and time periods that do not cancel out. For example, in the case where the placebo DiD includes only observations under treatment, unit $s r^\prime$ may have initiated treatment earlier than $s^\prime r^\prime$. Therefore, the placebo DiD includes a difference between the the ATT of the earlier timing group versus the ATT of the later timing group at two different time periods. Unless these effects are equal, the triple difference will fail to identify an ATT \textit{even if} the primary DiD is valid.

Finally, four terms in the decomposition include second-differences with an \textit{odd} number of treated observations -- that is, placebo DiDs that also act as primary DiDs elsewhere in the decomposition. A DiD comparison in one stratum $r$ is subtracted from a ``flipped" difference-in-difference in stratum $r^{\prime}$. Under constant treatment effects, this triple-difference eliminates the common parallel trends violation across the two strata and yields a term equivalent to twice the treatment effect. However, under effect heterogeneity the violation of parallel trends cannot be disentangled from differences in the treatment effect between stratum $r$ and $r^\prime$. As shown in the Appendix, these terms can also be re-written as ``double-counted" difference-in-differences providing an alternative interpretation of the triple-differences regression as an average over both 2x2x2 triple-differences and 2x2 difference-in-differences. 

The decomposition in Theorem \ref{thm:decomp} highlights how the magnitude of the bias in the conventional triple-differences regression under heterogeneous treatment effects depends both on the degree of effect heterogeneity and the extent of treatment staggering across strata. Under certain distributions of treatment, the fixed effects estimator may be comprised of more invalid triple-difference terms than valid ones. There are two useful implications of this result. First, even if there is no staggered adoption \textit{within} each stratum, the triple-differences regression estimator will still be biased under heterogeneous treatment effects if the specific units that enter treatment vary across each stratum. This implies that there are settings where, if conditional parallel trends holds for all strata (Assumption \ref{a:conditionalparalleltrends}), the triple-differences regression will be biased for an average of ATTs while the conventional two-way fixed effects regression will not be. Second, in settings where researchers have only a single placebo stratum and a single ``potentially treated" stratum, there are no invalid placebo terms. This provides something of a silver lining in Theorem \ref{thm:decomp} for the many triple-difference designs where researchers augment a conventional difference-in-difference using a pure placebo group. In these settings, researchers only need to be concerned about the already well-known issues with two-way fixed effects and invalid primary DiD terms as all of the placebo DiDs will exclusively incorporate observations under control.

As noted in the extensions of \textcite{borusyak2021revisiting}, an alternative estimation strategy using direct imputation of the counterfactual can easily address the issues of invalid comparisons in the triple-differences design. Instead of estimating the treatment effect in the \textit{same} model as the fixed-effects parameters, the imputation approach separates the task of modeling the potential outcomes under control and aggregating the imputed ATTs to obtain a causal quantity of interest. For the triple-differences setting, the counterfactual imputation estimator proceeds by estimating the triple-differences regression with three sets of fixed effects parameters among the observations observed under control ($D_{srt} = 0$). From this model, the method imputes the counterfactual under control for each unit/stratum/time period under treatment ($D_{srt} = 1$).
\begin{align*}
    \widehat{Y_{srt}(\infty)} = \hat{\alpha_{sr}^{(0)}} + \hat{\gamma_{st}^{(0)}} + \hat{\delta_{rt}^{(0)}}
\end{align*}

With the imputed counterfactuals, researchers can obtain an estimate of any stratum group-time treatment effect and any aggregation of the group-time treatment effects as in Equation \ref{eqn:aggregate}. For a more extensive discussion of the fixed effects counterfactual imputation estimator as applied to the difference-in-differences setting, see \textcite{liu2021practical}. 

Inference in the triple-difference imputation setting is complicated by the presence of potentially complex error correlation structures. In the conventional difference-in-difference setting, researchers typically assume error correlation within unit over time but independence across units. For the fixed effects imputation estimator, this facilitates inference via either the cluster bootstrap \parencite{liu2021practical} or via an asymptotic approximation \parencite{borusyak2021revisiting}. While a single cluster error correlation structure is plausible in some triple-difference settings, such as when observations are firms nested within affected states and industries, for many triple-difference applications errors likely exhibit a \textit{two-way} clustering structure where observations that share \textit{either} the same $s$ or the same $r$ are correlated. In particular, when observations are dyadic -- as in the ``gravity model of trade" setting examined in Section \ref{sec:application} -- it is implausible to assume that dyads that share a common member are independent despite the fact that many applied studies proceed with this assumption (\cite{aronow2015cluster}; \cite{carlson2021dyadic}). Rather, the clustering structure is ``two-way" in that errors across dyads are correlated on both the ``sender" dimension and the ``receiver" dimension and failing to account for this will tend to understate the true sampling variance. 

While asymptotic variance estimators that are robust to two-way clustering exist for the conventional regression model \parencite{cameron2011robust}, an extension to the imputation estimator in the style of \parencite{borusyak2021revisiting} for single clustering is beyond the scope of this paper. Instead, this paper suggests the use of an extension of the cluster bootstrap to the two-way clustering setting: the ``pigeonhole" bootstrap (\cite{owen2007pigeonhole}; \cite{owen2012bootstrapping}). This approach proceeds by resampling clusters along each of the clustering dimensions (e.g. resampling ``sender" units and ``receiver" units). The bootstrap weight assigned to each assigned to each dyad is the product of the bootstrap weights assigned to its sender and receiver. \textcite{owen2012bootstrapping} recommend a ``Bayesian" version of the pigeonhole bootstrap in the style of \textcite{rubin1981bayesian} in which weights on each cluster are drawn independently from an $\text{Exponential}(1)$ distribution. In the fixed-effects imputation setting this has a slight advantage over the conventional resampling bootstrap in that it does not assign a weight of $0$ to any dyad. This ensures that the number of fixed effects parameters remains the same when re-estimating the model in each bootstrap iteration. Recent theoretical work has shown asympottic consistency of the pigeonhole bootstrap under multi-way clustering (\cite{davezies2018asymptotic}, \cite{menzel2021bootstrap}) and \textcite{owen2012bootstrapping} note that it is generally conservative for the true sampling variance. To maintain comparability with existing applied work, I use a conventional one-way cluster bootstrap for the application in Section \ref{sec:application} but provide results under a two-way pigeonhole bootstrap in Appendix \ref{a:two-way}.

\section{Application: Gravity Models and the Effect of the WTO on Trade} \label{sec:application}

Despite the central role that the General Agreement on Tariffs and Trade (GATT) and its successor, the World Trade Organization (WTO) have played in the negotiated multilateral reduction of tariffs and other barriers to trade in the post-WWII era, there exists an extensive empirical debate over whether there is any clear evidence that membership in the GATT/WTO actually increases trade between states. \textcite{rose2004we} found little evidence that pairs of countries that were members of the GATT/WTO saw increased bilateral trade. \textcite{tomz2007we} and \textcite{goldstein2007institutions} responded by arguing that Rose failed to account for states that were participants in the GATT/WTO system and thus benefited from trade concessions despite not being full members. Re-estimating Rose's regressions using an alternative treatment indicator, these papers found evidence of a positive treatment effect. An extensive literature has followed which refined the original specifications and, for the most part, has found evidence of a GATT/WTO effect.\footnote{See \textcite{gil2016re} for a review. However, also note recent work by \textcite{esteve2020does} which provides some evidence for the original null under an alternative estimation strategy.}

There have been two primary econometric improvements over the original debate between Rose and Tomz, Goldstein and Rivers. The first is the use of correctly specified gravity models that incorporate terms that capture the unobserved heterogeneity common to sender and receiver that evolves over time. While \textcite{tomz2007we} and \textcite{goldstein2007institutions} used dyad and time fixed effects, the standard gravity model specification in the econometrics literature also includes an interaction between the time fixed effects and both sender and receiver fixed-effects to capture the ``multilateral resistance" factors that are common to an exporter and importer across all trading relationships \parencite{anderson2003gravity}. This approach was adopted in \textcite{subramanian2007wto} in a re-analysis of \textcite{rose2004we} and subsequent work has relied on a baseline regression specification that takes the form of the following log-linearized equation estimated via OLS that models trade between sender country $s$ and receiver country $r$ in time $t$ 
\begin{align*}
    \log(\text{Imports}_{srt}) = \mathbf{X}_{srt}^\prime \beta + \alpha_{sr} + \gamma_{st} + \delta_{rt} + \epsilon_{srt}
\end{align*}
where $\mathbf{X}_{srt}$ is a vector of covariates that vary across time and dyad and $\epsilon_{srt}$ is a mean-zero error term. As shown in this paper, with a single, binary $X_{srt}$, this corresponds exactly to the conventional triple-differences regression. This suggests that, implicitly, studies of policy effects on bilateral trade that implement a gravity equation as a baseline model for the outcome are relying on a form of Assumption \ref{a:constantviolation} to motivate their identification strategy.

The second econometric critique, which is largely beyond the scope of this paper, concerns the use of OLS to estimate a log-linearized gravity model. \textcite{silva2006log} note that under heteroskedasticity, OLS estimates for the structural parameters of the log-linearized gravity model parameters will be biased due to Jensen's inequality. As an alternative, they recommend instead estimating a regression on the raw trade levels using a multiplicative model for the conditional expectation function with parameters estimated via Poisson Pseudo-maximum likelihood (PPML) with robust standard errors. \textcite{wooldridge1999distribution} shows that this is consistent for the CEF under correct specification of the conditional mean function even in the presence of misspecification in the error distribution through the use of the Poisson. The PPML estimator has the added benefit of being able to incorporate zeroes which the log-linearized estimator must either ignore or transform using some arbitrary additive constant. Recently \textcite{esteve2020does} show that applying the PPML approach instead of the log-linearized model to estimate the effect of the GATT/WTO results in null findings in line with Rose's original results. 

While this application will focus on the log-linearized fixed effects gravity model estimated using OLS, the re-interpretation of the gravity model in terms of a triple-differences design also helps shed some light on this second critique from \textcite{silva2006log}. First, it suggests that zeroes in the trade flow data are not simply an estimation concern but rather a challenge to the underlying assumptions of the research design. As \textcite{ciani2018dif} note, the use of a log-linearized or Poisson model implicitly assumes that trends are multiplicative rather than additive on the level of the raw outcome. In the standard difference-in-differences setting this corresponds to an assumption that the \textit{ratios} of the potential outcomes under control rather than their \textit{differences} are equivalent between treated and control. In such a design, the presence of zeroes will, by construction, result in a violation of this assumption as the multiplicative trend relative to a baseline of zero is infinite. Therefore irrespective of the choice of estimator (log-linearized or Poisson PML), it may be appropriate to remove observations with zero trade flows to avoid including observations for which the identifying assumptions will not hold.

Second, it suggests that switching to a Poisson PML estimator alone will not resolve the problem of invalid triple-difference comparisons highlighted in this paper. While a full decomposition of the Poisson PML estimator is beyond the scope of this paper, recent work by \textcite{wooldridge2021two} for the difference-in-difference setting has noted that the conventional Poisson PML estimator with two-way fixed effects suffers from similar issues as TWFE OLS. However, simple corrections that allow for greater treatment effect heterogeneity (``extended" two-way fixed effects) can be applied to the PPML case as well as OLS. These estimators have an equivalent form to the imputation estimator in \textcite{borusyak2021revisiting} and therefore an imputation extension of the Poisson PML to the three-way fixed effects setting is plausible.

 This replication focuses on the differences between the conventional triple differences regression specification and the imputation estimator recommended by \textcite{borusyak2021revisiting} and \textcite{liu2021practical} as applied to the replication dataset from \textcite{goldstein2007institutions}. The original data consists of $381,656$ dyad-year observations from 1946 to 2004. I restrict the analysis to 1946-2003 as the published dataset exhibits a puzzling decline in the total number of observations from 2003 to 2004. While WTO membership is \textit{largely} a case of staggered adoption as there are very few states that revert from being members to non-members, there are three states which are considered early members that very quickly switch to being non-members: China, Lebanon and Syria. I drop the small number of observations where these states are considered treated (largely pre-1951). Among the WTO participants there are slightly more non-staggered cases as the definition of participation from \textcite{tomz2007we} includes a number of states that became de-facto participants after independence due to the participation of the GATT/WTO membership of their former colonial power. While many of these states remained in the system, a handful eventually dropped out as participants (and potentially rejoined in subsequent years). To retain the staggered adoption structure in the data, I drop observations from Vietnam pre-1956, Laos pre-1958, Guinea pre-1962, and Cambodia, Algeria and Yemen pre-1996. From the standpoint of the underlying design, pruning these observations is sensible as it is implausible to estimate treatment effects for these ``treated" periods absent any pre-treatment observations and these periods are also not valid controls for other units in the data. 

\begin{figure}[ht!]
    \centering
    \includegraphics[scale=.5]{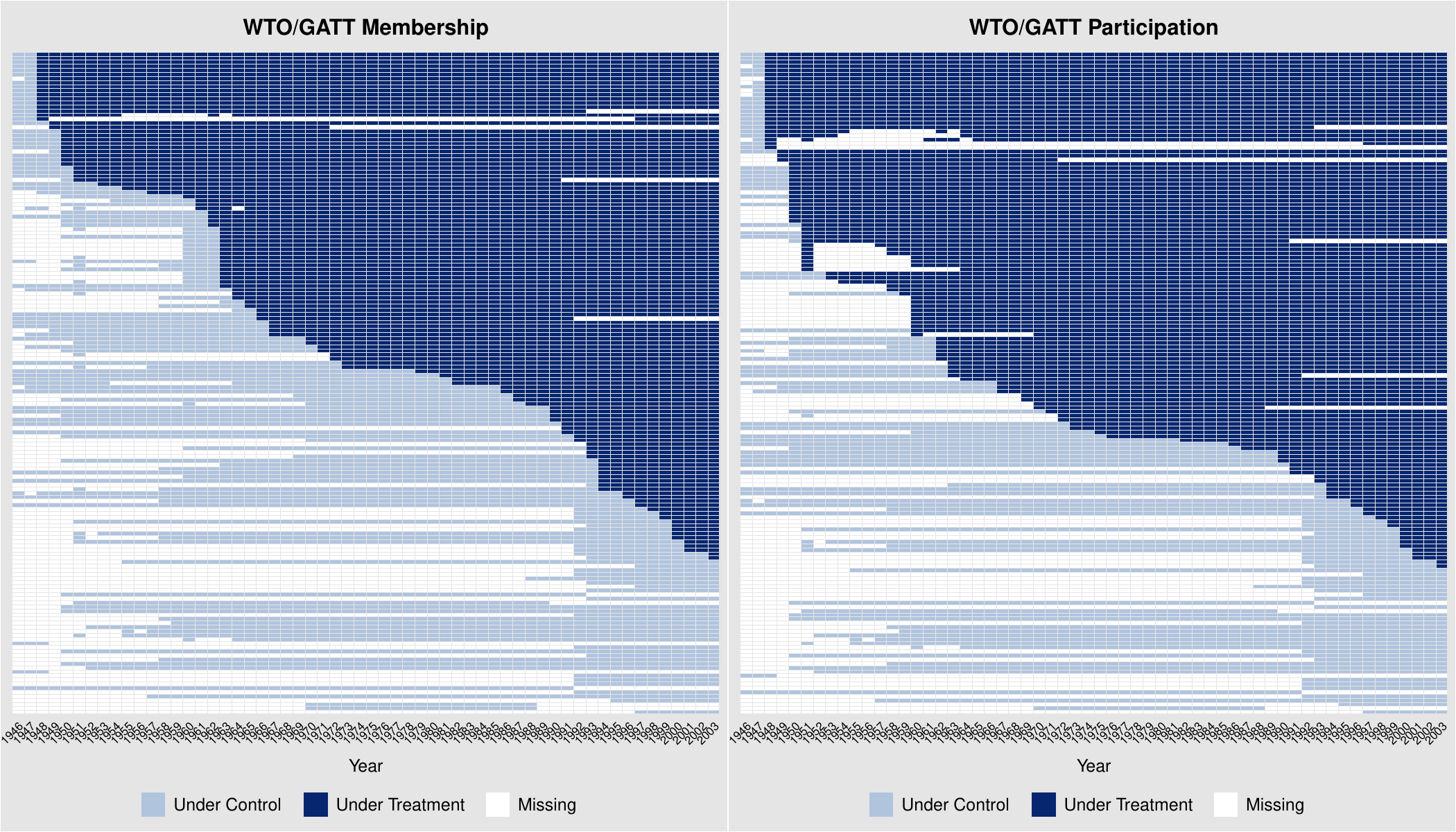}

    \raggedright \footnotesize \textit{Notes:} Data from \textcite{goldstein2007institutions}. 163 countries, 58 years.
    
    \caption{Monadic distribution of WTO membership and participation}
    \label{fig:distplot}
\end{figure}
 
After pre-processing, the dataset consists of $371,954$ dyad-years with $163$ unique countries and $58$. Following \textcite{tomz2007we}, I code two dyadic treatment variables: joint membership if both members of a dyad are a member of the WTO in year $t$ and joint participation if both members of a dyad are participants in year $t$. While a visualization of the treatment distribution at a dyadic level would be infeasible, I generate two treatment adoption plots \parencite{liu2021practical} at the monadic level to understand both the magnitude of the treatment staggering as well as the overall difference between membership and participation treatments. Figure \ref{fig:distplot} plots the two distributions with states ordered by year of membership/participation respectively. 

The results show both significant missingness in the data as dyads with zero trade flows were omitted and many states do not exist for the entire time period under analysis. They also highlight substantial staggering in both treatments. There are comparatively few states in later years that are not WTO/GATT members. The scarcity of ``pure control" observations also limits the ability of researchers to precisely estimate long-run effects, especially for the early members. Moreover, while it is clear that the distribution of participants is qualitatively different from members, this differences has consequences not only for treatment effect heterogeneity - as highlighted by \textcite{tomz2007we} and \textcite{goldstein2007institutions} - but also for the precision with which the treatment effects can be estimated. With many more treated units, there exist fewer control observations from which to impute. In general, I find the precision of estimates using joint participation as the treatment to be much greater than those using joint membership, especially when only imputing using control observations.

\begin{table}[ht!]

\caption{Estimated effects of dyadic WTO membership on log imports}\label{tab:main_results}
\centering
\begin{tabular}{l|cc|cc}
& \multicolumn{2}{c|}{Fixed-effects Regression} & \multicolumn{2}{c}{Imputation}\\
& Members & Participants & Members & Participants\\
\hline
Estimate & 	$0.0544$ & $0.0777$ &  $0.169$ & $0.176$\\
Std. Err & $(0.0250)$ & $(0.0281)$ & $(0.0629)$ & $(0.0718)$ \\
95\% CI & $[0.00529, 0.103]$ & $[0.02256, 0.133]$ & $[0.0458, 0.292]$ & $[0.0354, 0.317]$\\
\hline
\end{tabular}

\raggedright \footnotesize \textit{Notes:} $371,954$ dyads, $163$ countries, $58$ time-periods. Standard errors are cluster-bootstrapped clustered on dyad. 100 bootstrap iterations.

\end{table}

I estimate the log-linearized three-way fixed effects regression model with log-imports as the outcome. Table \ref{tab:main_results} reports the estimated average treatment effects from the ``static" specification. For the conventional ``fixed-effects regression" I report the estimated coefficient on the treatment indicator (either joint membership or joint participation) while for the imputation results, I impute the counterfactual for each treated dyad using the three-way fixed effects model fit to the controls and average over the difference between observed and imputed trade uniformly across all dyads and time periods. Standard errors are estimated using the Bayesian dyadic bootstrap, using a random weight to re-weighting each dyad while preserving intra-dyad correlations over time. This is consistent with the standard approach in much of the WTO/GATT effects literature which only clusters on dyad. However, this likely under-estimates the true sampling variance. Appendix \ref{a:two-way} presents these results with standard errors obtained via the two-way pigeonhole bootstrap and suggests that few of these estimated effects are actually statistically significant.

Two features from the static results are readily apparent, the fixed-effects regression estimates are more attenuated towards zero but the standard errors are much smaller. This is a consequence of the use of \textit{more} observations but also of the use of potentially \textit{invalid} triple-difference comparisons. As noted in the DiD case by simulations in \textcite{baker2022much}, when treatment effects accumulate over time, two-way fixed effects estimators will tend to under-estimate the ATT. As it is unlikely that the effects of trade liberalization appear instantaneously over a single year, it is sensible to expect the effects of the WTO to be cumulative. Eliminating these invalid comparisons via the imputation estimator shifts the estimated ATT upwards by a factor of 3. Notably, this shift is considerably greater than the difference between the estimates for members versus participants, suggesting that questions of model specification and design are much more salient than distinctions in how the treatment is coded. But the imputation estimators come at a cost -- a substantial increase in the variance resulting from the use of fewer relevant observations to impute the counterfactual and from an alternative weighting of the individual treatment effects.

\begin{figure}[ht!]
    \centering
    \includegraphics[scale=.45]{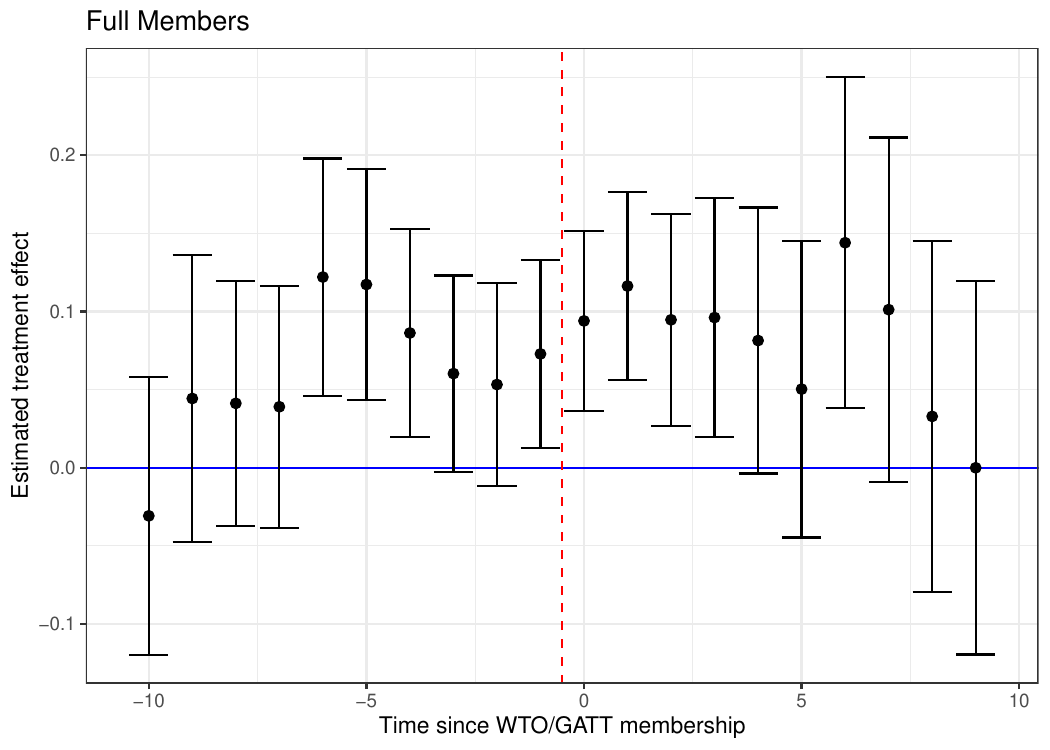}\includegraphics[scale=.45]{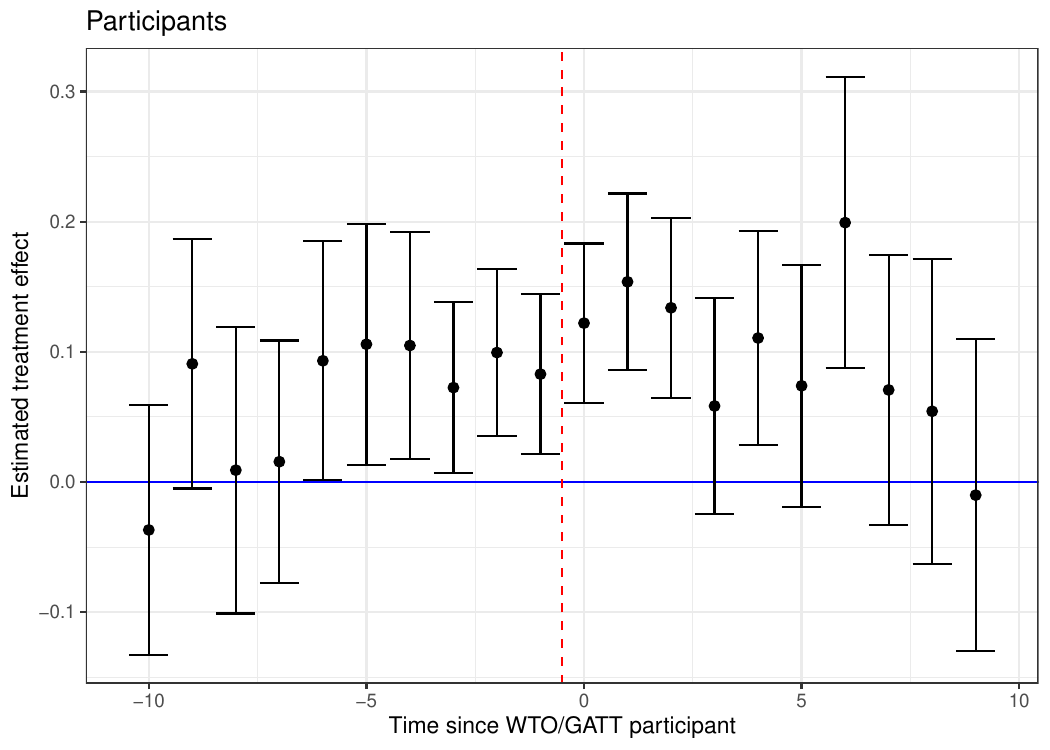}

    \raggedright \footnotesize \textit{Notes:} $371,954$ dyads, $163$ countries, $58$ time-periods. Time = $0$ denotes the first period under treatment. Post-treatment estimates (0 to 10) averaged from imputed effects across all treated units $t$ periods after treatment adoption. Pre-treatment estimates (-10 to -1) are held-out placebo tests. Bars denote 95\% cluster-bootstrapped CIs clustered on dyad. 100 bootstrap iterations.
    
    \caption{Estimated effects of dyadic WTO membership on log imports - effects over time up to 10 years post-treatment}
    \label{fig:eventplot}
\end{figure}

The imputation approach also facilitates the creation of ``event-study" plots for the triple-differences setting. This allows researchers to not only assess whether the treatment effect is heterogeneous over time, but also to conduct ``placebo tests" for violations of the identifying assumptions by leaving out each pre-treatment period and imputing the counterfactual from the remaining observations. Figure \ref{fig:eventplot} plots the estimated average treatment effect using the imputation approach for up to 10 periods post-treatment for both the joint membership and the joint participation treatment. It also implements the placebo test method of \textcite{liu2021practical} for up to 10 periods pre-treatment. For each pre-treatment lag, the placebo test assumes that all units actually initiated treatment that many periods earlier and estimates the imputation model using only never-treated dyads and the time periods prior to the lag. Using this model, it generates a counterfactual prediction for these pre-treatment periods. If the identifying assumptions hold, the difference between the observed outcome and the imputed outcome should be zero. Conversely, statistically significant pre-trends effects are evidence that the identifying assumptions may be invalid and that there may be a violation of Assumption \ref{a:constantviolation}.

Focusing on the first 10 years before and after treatment, I find some evidence that the constant violation of parallel trends assumption is invalid in the case of WTO/GATT membership and participation. While the estimated ATTs one to four years post joint-membership are positive and statistically significant, the magnitudes are comparable to the placebo estimates for the period 1 to 5 years \textit{prior} to entry. Analogous results appear for joint participation -- the placebo effects are all positive and statistically significant. These placebos provide some evidence that states which enter the WTO are likely altering their behavior prior to both membership and participation \textit{specifically} with respect to other WTO members. As a result, evidence for a short-run effect is particularly weak given the likelihood of anticipation in the design.

\begin{figure}[ht!]
    \centering
    \includegraphics[scale=.45]{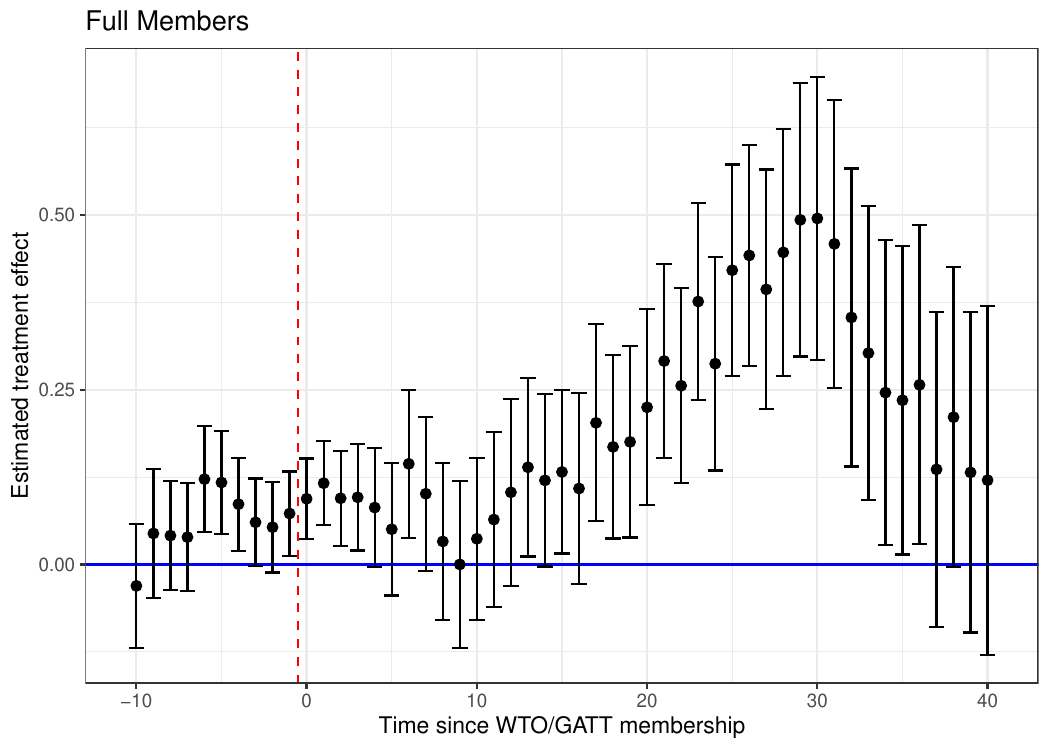} \includegraphics[scale=.45]{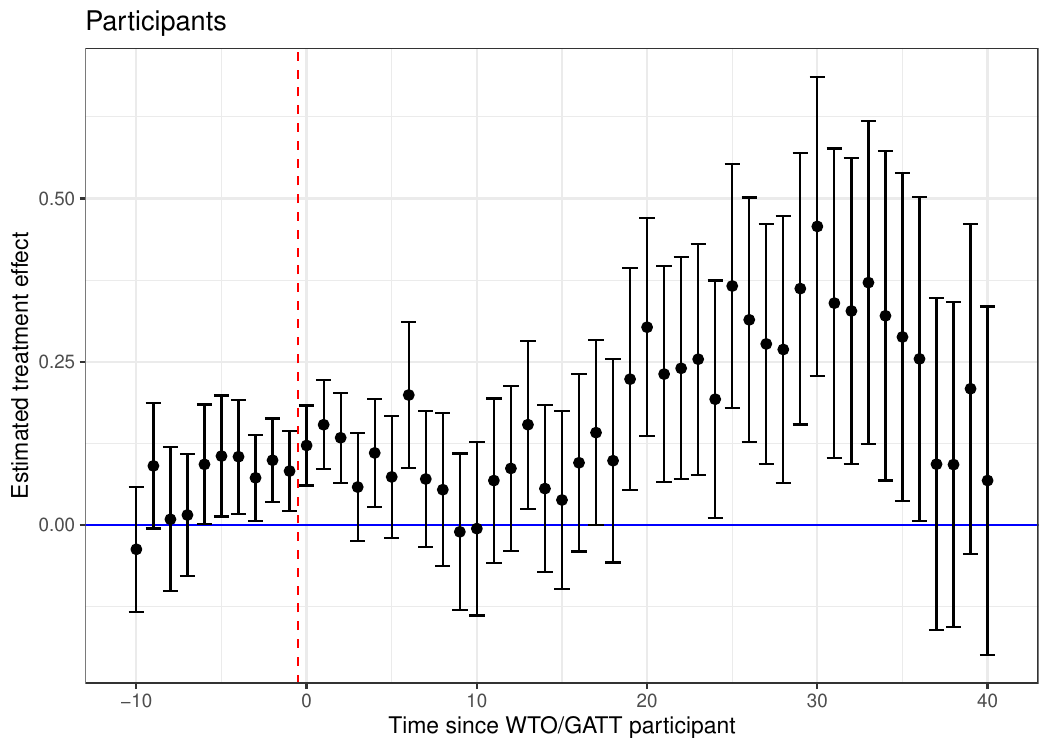}
    
    \raggedright \footnotesize \textit{Notes:} $371,954$ dyads, $163$ countries, $58$ time-periods. Time = $0$ denotes the first period under treatment. Post-treatment estimates (0 to 40) averaged from imputed effects across all treated units $t$ periods after treatment adoption. Pre-treatment estimates (-10 to -1) are held-out placebo tests. Bars denote 95\% cluster-bootstrapped CIs clustered on dyad. 100 bootstrap iterations.
    
    \caption{Estimated effects of dyadic WTO membership on log imports - effects over time up to 40 years post-treatment}
    \label{fig:eventplot_long}
\end{figure}

However, expanding the number of post-treatment periods being considered, does suggest \textit{some} long-run impact of WTO/GATT membership. In Figure \ref{fig:eventplot_long}, I extend the number of post-treatment years considered to 40. I find sizeable effects on log imports 20-30 years post-membership with magnitudes much larger than the placebo estimates. Similar patterns appear for participants, suggesting that there may be some evidence for an \textit{incremental} effect of WTO/GATT membership over time as opposed to an instantaneous effect at the time of membership. However, these findings come with a few caveats. First, as shown in Appendix \ref{a:two-way}, these results are not robust to the use of two-way clustered standard errors. Second, these ``time-since-treatment" effects are not necessarily directly comparable to one another as the composition of treatment cohorts differs under staggered adoption -- for example, there are some treated states that only have 3 or 4 post-treatment periods and thus disappear from estimates of the ATT. For the long event study plot, effect estimates 40 years out can only incorporate treatment groups that entered into the WTO before 1963. The consequence is that inter-temporal heterogeneity in treatment effects may be conflated with variation across treatment timing groups. Changes in which states comprise each treatment timing group could explain the sudden drop in the estimated effect between 30 and 40 years post-treatment.

Overall, the replication results suggest that \textit{any} conclusions about the effect of the WTO/GATT on trade obtained from conventional gravity model regressions should be taken with a grain of salt. Even after addressing the biases of the static gravity model using an imputation estimator, there is strong evidence from the statistically significant placebo estimates that the underlying identifying assumptions  are violated. The replication findings should lead trade scholars to exercise some caution in estimating causal effects from structural models -- the gravity model alone does not guarantee identification, it is only a model for the control counterfactual. Whether this model is valid for causal identification is not just a question of trade theory, it is also a question about the treatment assignment mechanism. In fact, the gravity model imposes a very specific set of assumptions on the control potential outcomes that imply an underlying \textit{triple-differences} research design. While researchers should continue to use the gravity model in cases where these implied identification assumptions are plausible, this paper cautions against estimating the treatment effect and the gravity model parameters simultaneously in a single regression as is standard in the empirical trade literature. Instead, researchers should consider using the gravity model specification as a model for the potential outcomes under control. Then, after fitting this model to control observations, researchers can \textit{impute} the counterfactuals for units under treatment. This imputation approach also facilitates straightforward placebo tests to diagnose violations of the identifying assumptions behind the gravity model -- a potentially very useful technique that has not yet been widely adopted in studies of international trade.

\section{Discussion}

Triple-differences designs are growing in popularity among applied researchers, but a formal discussion of their identifying assumptions has been largely absent from the literature until recently with \textcite{olden2020triple} developing theory for identification in the basic 2x2x2 design. However, for researchers working with datasets with staggered adoption and multiple placebo strata, guidance has been, until recently, extremely limited and many empirical papers still rely on the ``three-way" fixed effects regression to estimate treatment effects.

This paper develops a theoretical framework for understanding the identifying assumptions of triple-differences under staggered adoption and demonstrates via a decomposition in the style of \textcite{goodman2021difference} why the commonly-used regression triple differences estimator does not identify an average of group-time ATTs under unrestricted treatment effect heterogeneity. It highlights how the triple-differences regression can be expressed as an average over 2x2x2 triple-difference terms that contain a ``primary" and a ``placebo" DiD. Unless treatment effects are constant across all units, time periods and strata, many of these placebo DiDs contain treated observations and will fail to adjust for the bias in the primary DiD even if the triple-differences identifying assumptions hold. 

Luckily for researchers, alternatives to the triple-differences regression estimator already exist. \textcite{borusyak2021revisiting} note that an estimator based on direct imputation of the counterfactual can be straightforwardly implemented for triple-differences designs using the same three-way fixed effects structure. Fitting the triple-differences regression among \textit{only the control} observations and imputing the counterfactuals under treatment addresses both of the problems of invalid primary and placebo DiDs shown in the decomposition. Moreover, the decomposition highlights how most of the problems of triple differences regression occur when there exist multiple strata that contain both primary and placebo DiDs. If, as is often the case, a researcher implements a triple-difference by appending a stratum known to \textit{never} receive treatment to a conventional staggered adoption design, no invalid placebos appear in the regression estimator.

There are a number of directions for future work in this area. First, the decomposition focuses on the ``static" fixed effects regression, but researchers often also estimate ``dynamic" regressions with indicators for treatment leads and lags to allow for some treatment effect heterogeneity over time. As such an extension of the results in \textcite{sun2021estimating} to the triple-differences setting would clarify the extent to which these estimates are contaminated by comparisons that are invalid under unrestricted effect heterogeneity. Second, the present decomposition considers a balanced panel with no missing data while many applied settings -- including the trade example in this paper -- have substantial amounts of missingness in both treatment and outcome. This affects the weights assigned to each comparison and each ATT in the triple differences regression. Finally, this paper presently only evaluates the direct counterfactual imputation estimator from \textcite{borusyak2021revisiting}. However, other heterogeneity-robust estimators from the differences-in-differences setting, such as the doubly-robust estimator in \textcite{callaway2021difference} which incorporates both a treatment \textit{and} an outcome model, could also be extended to the ``many stratum" triple-differences setting.\footnote{With only a single primary and a single placebo stratum, the existing difference-in-differences methods are straightforward to adapt simply by redefining the outcome of interest as the difference between the outcomes in the primary and placebo strata. However, with many placebo strata and differentially staggered treatment, the set of valid placebo comparisons will depend on the particular stratum-specific group-time ATT.}

\clearpage

\printbibliography

\clearpage

\begin{appendices}

\section{Proofs}

\subsection{Proof of Proposition \ref{prop:IDparallel}}

This proof is a special case of the more general doubly-robust result from \textcite{callaway2021difference} that allows for parallel trends to hold conditional on $X$.

Start by writing the potential outcomes under Consistency/SUTVA (Assumption \ref{a:sutva-aggregate}).
\begin{align*}
\mathbb{E}[Y_{srt} - Y_{srt^*} | G_{sr} = g] -  \mathbb{E}[Y_{srt} - Y_{srt^*} | G_{sr} = g^{\prime}] &= \mathbb{E}[Y_{srt}(g) - Y_{srt^*}(g) | G_{sr} = g] \\ &-  \mathbb{E}[Y_{srt}(g^{\prime}) - Y_{srt^*}(g^{\prime}) | G_{sr} = g^{\prime}]
\end{align*}

Under no anticipation (Assumption \ref{a:noanticipation}), since $t^* < g$, $t < g^{\prime}$, and by extension $t^* < g^{\prime}$
\begin{align*}
\mathbb{E}[Y_{srt} - Y_{srt^*} | G_{sr} = g] -  \mathbb{E}[Y_{srt} - Y_{srt^*} | G_{sr} = g^{\prime}] &=  \mathbb{E}[Y_{srt}(g) - Y_{srt^*}(\infty) | G_{sr} = g] \\ &-  \mathbb{E}[Y_{srt}(\infty) - Y_{srt^*}(\infty) | G_{sr} = g^{\prime}]
\end{align*}

Under conditional parallel trends (Assumption \ref{a:conditionalparalleltrends}), we have
\begin{align*}
     E[Y_{srt^{*}}(\infty) | G_{sr} = g] + \mathbb{E}[Y_{srt}(\infty) - Y_{srt^{*}}(\infty) | G_{sr} = g^{\prime}] = \mathbb{E}[Y_{srt}(\infty) | G_{sr} = g]
\end{align*}

Substituting, into the above expression, we have the definition of the $\text{ATT}_r(g,t)$
\begin{align*}
\mathbb{E}[Y_{srt} - Y_{srt^*} | G_{sr} = g] -  \mathbb{E}[Y_{srt} - Y_{srt^*} | G_{sr} = g^{\prime}] &=  \mathbb{E}[Y_{srt}(g) - Y_{srt}(\infty) | G_{sr} = g]\\
&= \text{ATT}_r(g,t)
\end{align*}

\subsection{Proof of Proposition \ref{prop:IDtripdiff}}

Start with the first difference-in-difference. Under Consistency/SUTVA (Assumption \ref{a:sutva-aggregate}).
\begin{align*}
\mathbb{E}[Y_{srt} - Y_{srt^*} | G_{sr} = g] -  \mathbb{E}[Y_{srt} - Y_{srt^*} | G_{sr} = g^{\prime}] &= \mathbb{E}[Y_{srt}(g) - Y_{srt^*}(g) | G_{sr} = g] \\ &-  \mathbb{E}[Y_{srt}(g^{\prime}) - Y_{srt^*}(g^{\prime}) | G_{sr} = g^{\prime}] 
\end{align*}

Under no anticipation (Assumption \ref{a:noanticipation}), since $t^* < g$, $t < g^{\prime}$, and by extension $t^* < g^{\prime}$
\begin{align*}
\mathbb{E}[Y_{srt} - Y_{srt^*} | G_{sr} = g] -  \mathbb{E}[Y_{srt} - Y_{srt^*} | G_{sr} = g^{\prime}] &= \mathbb{E}[Y_{srt}(g) - Y_{srt^*}(\infty) | G_{sr} = g] \\ &-  \mathbb{E}[Y_{srt}(\infty) - Y_{srt^*}(\infty) | G_{sr} = g^{\prime}] 
\end{align*}

Under the constant violation of parallel trends assumption (Assumption \ref{a:constantviolation})
\begin{align*}
    \mathbb{E}[Y_{srt^{*}}(\infty) | G_{sr} = g] + \mathbb{E}[Y_{srt}(\infty) - Y_{srt^{*}}(\infty) | G_{sr} = g^{\prime}] &=  \mathbb{E}[Y_{srt}(\infty) | G_{sr} = g] \\ &- \mathbb{E}[Y_{sr^{\prime}t}(\infty) - Y_{sr^{\prime}t^{*}}(\infty) | G_{sr} = g] \\ &+ \mathbb{E}[Y_{sr^{\prime}t}(\infty) - Y_{sr^{\prime}t^{*}}(\infty) | G_{sr} = g^{\prime}]
\end{align*}

Substituting into the first DiD expression, we have the $\text{ATT}_r(g,t)$ plus a bias term
\begin{align*}
\mathbb{E}[Y_{srt} - Y_{srt^*} | G_{sr} = g] -  \mathbb{E}[Y_{srt} - Y_{srt^*} | G_{sr} = g^{\prime}] &=  \mathbb{E}[Y_{srt}(g) - Y_{srt}(\infty) | G_{sr} = g] \\ &+ \mathbb{E}[Y_{sr^{\prime}t}(\infty) - Y_{sr^{\prime}t^{*}}(\infty) | G_{sr} = g] \\ &- \mathbb{E}[Y_{sr^{\prime}t}(\infty) - Y_{sr^{\prime}t^{*}}(\infty) | G_{sr} = g^{\prime}]
\end{align*}

Next, for the second-difference in-difference, under Consistency/SUTVA (Assumption \ref{a:sutva-aggregate}) along with no anticipation (Assumption \ref{a:noanticipation}), since $G_{sr^{\prime}} > t$ and by extension, $G_{sr^{\prime}} > t^*$
\small
\begin{align*}
\mathbb{E}[Y_{sr^{\prime}t} - Y_{sr^{\prime}t^*} | G_{sr} = g, G_{sr^{\prime}} > t ] -  \mathbb{E}[Y_{sr^{\prime}t} - Y_{sr^{\prime}t^*} | G_{sr} = g^{\prime},  G_{sr^{\prime}} > t] &= \mathbb{E}[Y_{sr^{\prime}t}(\infty) - Y_{sr^{\prime}t^*}(\infty) | G_{sr} = g, G_{sr^{\prime}} > t ] \\&-  \mathbb{E}[Y_{sr^{\prime}t}(\infty) - Y_{sr^{\prime}t^*}(\infty) | G_{sr} = g^{\prime},  G_{sr^{\prime}} > t] 
\end{align*}
\normalsize

Since we assume constant violation of parallel trends holds for all $r$, $r^\prime$, $r \neq r^\prime$
\small
\begin{align*}
&\mathbb{E}[Y_{sr^{\prime}t}(\infty) - Y_{sr^{\prime}t^*}(\infty) | G_{sr} = g, G_{sr^{\prime}} > t ] - \mathbb{E}[Y_{sr^{\prime}t}(\infty) - Y_{sr^{\prime}t^*}(\infty) | G_{sr} = g^{\prime},  G_{sr^{\prime}} > t]  = \\ &\mathbb{E}[Y_{sr^{\prime}t}(\infty) - Y_{sr^{\prime}t^*}(\infty) | G_{sr} = g] -  \mathbb{E}[Y_{sr^{\prime}t}(\infty) - Y_{sr^{\prime}t^*}(\infty) | G_{sr} = g^{\prime}] 
\end{align*}
\normalsize

Therefore, the second difference-in-difference equals the bias term from the first difference and we have 
\begin{align*}
    \text{ATT}_r(g,t) = \bigg\{\mathbb{E}[Y_{srt} - Y_{srt^*} | G_{sr} = g] -  \mathbb{E}[Y_{srt} - Y_{srt^*} | G_{sr} = g^{\prime}]\bigg\} -\\
    \bigg\{\mathbb{E}[Y_{sr^{\prime}t} - Y_{sr^{\prime}t^*} | G_{sr} = g, G_{sr^{\prime}} > t ] -  \mathbb{E}[Y_{sr^{\prime}t} - Y_{sr^{\prime}t^*} | G_{sr} = g^{\prime},  G_{sr^{\prime}} > t]\bigg\}
\end{align*}

\subsection{Proof of Lemma \ref{lem:ols}}

By Frisch-Waugh-Lovell, we can write $\hat{\tau}$ as the regression of $Y_{srt}$ on $\tilde{D}_{srt}$, the residual from an OLS regression of $D_{srt}$ on the unit and grouping-time fixed effects.
\begin{align*}
    \tilde{D}_{srt} = D_{srt} - \hat{D}_{srt}
\end{align*}

where 
\begin{align*}
    \hat{D}_{srt} = \hat{\tilde{\alpha}}_{sr} + \hat{\tilde{\gamma}}_{st} + \hat{\tilde{\delta}}_{rt}
\end{align*}

and the coefficients are solutions to the OLS minimization problem.
\begin{align*}
    \hat{\tilde{\alpha}}, \hat{\tilde{\gamma}}, \hat{\tilde{\delta}} = \argmin_{\tilde{\alpha}, \tilde{\gamma}, \tilde{\delta}} \sum_{s=1}^S \sum_{r=1}^R \sum_{t=1}^T (D_{srt} -  \tilde{\alpha}_{sr} - \tilde{\gamma}_{st} - \tilde{\delta}_{rt})^2
\end{align*}

Denote the means
\begin{align*}
    \bar{D}_{sr} &= \frac{1}{T} \sum_{t^{\prime}=1}^T D_{srt^{\prime}} &
    \bar{D}_{st} &= \frac{1}{R} \sum_{r^{\prime}=1}^R D_{sr^{\prime}t} &
    \bar{D}_{rt} &= \frac{1}{S} \sum_{s^{\prime}=1}^S D_{s^{\prime}rt}\\
    \bar{\bar{D}}_{s} &= \frac{1}{RT} \sum_{t^{\prime}=1}^T \sum_{r^{\prime}=1}^R D_{sr^{\prime}t^{\prime}} &
    \bar{\bar{D}}_{r} &= \frac{1}{ST} \sum_{t^{\prime}=1}^T \sum_{s^{\prime}=1}^S D_{s^{\prime}rt^{\prime}} &
    \bar{\bar{D}}_{t} &= \frac{1}{SR} \sum_{s^{\prime}=1}^S \sum_{r^{\prime}=1}^R D_{s^{\prime}rt^{\prime}}\\
    \bar{\bar{\bar{D}}} &= \frac{1}{SRT} \sum_{s^{\prime}=1}^S \sum_{r^{\prime}=1}^R \sum_{t^{\prime}=1}^T D_{s^{\prime}r^{\prime}t^{\prime}}\\
\end{align*}

First-order conditions
\begin{align*}
    \hat{\tilde{\alpha}}_{sr} &= \frac{1}{T} \sum_{t=1}^T D_{srt} - \frac{1}{T} \sum_{t=1}^T \hat{\tilde{\gamma}}_{st} - \frac{1}{T} \sum_{t=1}^T \hat{\tilde{\delta}}_{rt}\\
    \hat{\tilde{\gamma}}_{st} &= \frac{1}{R} \sum_{r=1}^R D_{srt} - \frac{1}{R} \sum_{r=1}^R \hat{\tilde{\alpha}}_{sr} - \frac{1}{R} \sum_{r=1}^R \hat{\tilde{\delta}}_{rt}\\
    \hat{\tilde{\delta}}_{rt} &= \frac{1}{S} \sum_{s=1}^S D_{srt} - \frac{1}{S} \sum_{s=1}^S \hat{\tilde{\alpha}}_{sr} - \frac{1}{S} \sum_{s=1}^S \hat{\tilde{\gamma}}_{st}
\end{align*}

Writing the predicted value $\hat{D}_{srt} = \hat{\tilde{\alpha_{sr}}} +  \hat{\tilde{\gamma_{st}}} + \hat{\tilde{\delta_{rt}}}$
\begin{align*}
    \hat{D}_{srt} &= \frac{1}{T} \sum_{t^{\prime}=1}^T D_{srt^{\prime}} +  \frac{1}{R} \sum_{r=1}^R D_{sr^{\prime}t} + \frac{1}{S} \sum_{s=1}^S D_{s^{\prime}rt}
    \\&  - \frac{1}{T} \sum_{t^{\prime}=1}^T \hat{\tilde{\gamma}}_{st^{\prime}} - \frac{1}{T} \sum_{t^{\prime}=1}^T \hat{\tilde{\delta}}_{rt^{\prime}}
    \\& - \frac{1}{R} \sum_{r^{\prime}=1}^R \hat{\tilde{\alpha}}_{sr^{\prime}} - \frac{1}{R} \sum_{r^{\prime}=1}^R \hat{\tilde{\delta}}_{r^{\prime}t}
    \\& - \frac{1}{S} \sum_{s^{\prime}=1}^S \hat{\tilde{\alpha}}_{s^{\prime}r} - \frac{1}{S} \sum_{s^{\prime}=1}^S \hat{\tilde{\gamma}}_{s^{\prime}t}
\end{align*}

Substituting
\begin{align*}
    \hat{D}_{srt} &= \frac{1}{T} \sum_{t^{\prime}=1}^T D_{srt^{\prime}} +  \frac{1}{R} \sum_{r=1}^R D_{sr^{\prime}t} + \frac{1}{S} \sum_{s=1}^S D_{s^{\prime}rt}
    \\&  - \frac{1}{RT} \sum_{t^{\prime}=1}^T \sum_{r^{\prime}=1}^R D_{sr^{\prime}t^{\prime}} + \frac{1}{RT} \sum_{t^{\prime}=1}^T \sum_{r^{\prime}=1}^R \hat{\tilde{\alpha}}_{sr^{\prime}} + \frac{1}{RT} \sum_{t^{\prime}=1}^T \sum_{r^{\prime}=1}^R \hat{\delta}_{r^{\prime}t^{\prime}}
    \\& - \frac{1}{ST} \sum_{t^{\prime}=1}^T \sum_{s^{\prime} = 1}^S D_{s^{\prime}rt^{\prime}} + \frac{1}{ST} \sum_{t^{\prime}=1}^T \sum_{s^{\prime}=1}^S \hat{\tilde{\alpha}}_{s^{\prime}r} + \frac{1}{ST} \sum_{t^{\prime}=1}^T \sum_{s^{\prime}=1}^S \hat{\tilde{\gamma}}_{s^{\prime}t^{\prime}}
    \\& - \frac{1}{SR} \sum_{s^{\prime}=1}^S \sum_{r^{\prime}=1}^R D_{s^{\prime} r^{\prime} t} + \frac{1}{SR} \sum_{s^{\prime}=1}^S \sum_{r^{\prime}=1}^R \hat{\tilde{\alpha}}_{s^{\prime}r^{\prime}} + \frac{1}{SR} \sum_{s^{\prime}=1}^S \sum_{r^{\prime}=1}^R \hat{\tilde{\delta}}_{r^{\prime}t} 
    \\& - \frac{1}{R} \sum_{r^{\prime}=1}^R \hat{\tilde{\alpha}}_{sr^{\prime}} - \frac{1}{R} \sum_{r^{\prime}=1}^R \hat{\tilde{\delta}}_{r^{\prime}t} -  \frac{1}{S} \sum_{s^{\prime}=1}^S \hat{\tilde{\alpha}}_{s^{\prime}r}
\end{align*}

Re-arranging + cancelling
\begin{align*}
    \hat{D}_{srt} &= \frac{1}{T} \sum_{t^{\prime}=1}^T D_{srt^{\prime}} +  \frac{1}{R} \sum_{r=1}^R D_{sr^{\prime}t} + \frac{1}{S} \sum_{s=1}^S D_{s^{\prime}rt}
    \\&  - \frac{1}{RT} \sum_{t^{\prime}=1}^T \sum_{r^{\prime}=1}^R D_{sr^{\prime}t^{\prime}} - \frac{1}{ST} \sum_{t^{\prime}=1}^T \sum_{s^{\prime} = 1}^S D_{s^{\prime}rt^{\prime}} - \frac{1}{SR} \sum_{s^{\prime}=1}^S \sum_{r^{\prime}=1}^R D_{s^{\prime} r^{\prime} t} 
    \\& + \frac{1}{RT} \sum_{t^{\prime}=1}^T \sum_{r^{\prime}=1}^R \hat{\delta}_{r^{\prime}t^{\prime}}
     + \frac{1}{ST} \sum_{t^{\prime}=1}^T \sum_{s^{\prime}=1}^S \hat{\tilde{\gamma}}_{s^{\prime}t^{\prime}}  + \frac{1}{SR} \sum_{s^{\prime}=1}^S \sum_{r^{\prime}=1}^R \hat{\tilde{\alpha}}_{s^{\prime}r^{\prime}} 
\end{align*}

Substituting again
\begin{align*}
    \hat{D}_{srt} &= \frac{1}{T} \sum_{t^{\prime}=1}^T D_{srt^{\prime}} +  \frac{1}{R} \sum_{r=1}^R D_{sr^{\prime}t} + \frac{1}{S} \sum_{s=1}^S D_{s^{\prime}rt}
    \\&  - \frac{1}{RT} \sum_{t^{\prime}=1}^T \sum_{r^{\prime}=1}^R D_{sr^{\prime}t^{\prime}} - \frac{1}{ST} \sum_{t^{\prime}=1}^T \sum_{s^{\prime} = 1}^S D_{s^{\prime}rt^{\prime}} - \frac{1}{SR} \sum_{s^{\prime}=1}^S \sum_{r^{\prime}=1}^R D_{s^{\prime} r^{\prime} t} 
    \\& + \frac{1}{RT} \sum_{t^{\prime}=1}^T \sum_{r^{\prime}=1}^R \hat{\delta}_{r^{\prime}t^{\prime}}
     + \frac{1}{ST} \sum_{t^{\prime}=1}^T \sum_{s^{\prime}=1}^S \hat{\tilde{\gamma}}_{s^{\prime}t^{\prime}} 
    \\& + \frac{1}{SRT} \sum_{s^{\prime}=1}^S \sum_{r^{\prime}=1}^R \sum_{t^{\prime} = 1}^T D_{s^{\prime}r^{\prime}t^{\prime}} -  \frac{1}{SRT} \sum_{s^{\prime}=1}^S \sum_{r^{\prime}=1}^R \sum_{t^{\prime} = 1}^T \hat{\tilde{\gamma}}_{s^{\prime}t^{\prime}} -  \frac{1}{SRT} \sum_{s^{\prime}=1}^S \sum_{r^{\prime}=1}^R \sum_{t^{\prime} = 1}^T \hat{\tilde{\delta}}_{r^{\prime}t^{\prime}} 
\end{align*}

Cancelling again yields the residual $\tilde{D}_{srt}$ in terms of averages of $D_{srt}$ across dimensions $s$, $r$ and $t$.
\begin{align*}
    \tilde{D}_{srt} &= D_{srt} - \bar{D}_{sr} - \bar{D}_{st} - \bar{D}_{rt} + \bar{\bar{D}}_s + \bar{\bar{D}}_r + \bar{\bar{D}}_t - \bar{\bar{\bar{D}}}
\end{align*}

Returning to the expression for $\hat{\tau}$ yields
\begin{align*}
\hat{\tau} = \frac{\sum_{s=1}^S \sum_{r=1}^R \sum_{t=1}^T Y_{srt} \bigg(D_{srt} - \bar{D}_{sr} - \bar{D}_{st} - \bar{D}_{rt} + \bar{\bar{D}}_s + \bar{\bar{D}}_r + \bar{\bar{D}}_t - \bar{\bar{\bar{D}}}\bigg)}{\sum_{s=1}^S \sum_{r=1}^R \sum_{t=1}^T \bigg(\tilde{D}_{srt}\bigg)^2}
\end{align*}

Swapping indices in the numerator
\begin{align*}
\hat{\tau} = \frac{\sum_{s=1}^S \sum_{r=1}^R \sum_{t=1}^T D_{srt} \bigg(Y_{srt} - \bar{Y}_{sr} - \bar{Y}_{st} - \bar{Y}_{rt} + \bar{\bar{Y}}_s + \bar{\bar{Y}}_r + \bar{\bar{Y}}_t - \bar{\bar{\bar{Y}}}\bigg)}{\sum_{s=1}^S \sum_{r=1}^R \sum_{t=1}^T \bigg(\tilde{D}_{srt}\bigg)^2}
\end{align*}

With a binary treatment, we can re-write the numerator as a sum over $t$ with $D_{srt} = 1$.
\begin{align*}
   \hat{\tau} &= \frac{\sum_{s=1}^S \sum_{r=1}^R \sum_{t:D_{srt} = 1} Y_{srt} - \bar{Y}_{sr} - \bar{Y}_{st} - \bar{Y}_{rt} + \bar{\bar{Y}}_s + \bar{\bar{Y}}_r + \bar{\bar{Y}}_t - \bar{\bar{\bar{Y}}}}{\sum_{s=1}^S \sum_{r=1}^R \sum_{t=1}^T \bigg(\tilde{D}_{srt}\bigg)^2}
\end{align*}

\subsection{Proof of Theorem \ref{thm:decomp}}

Start from Lemma \ref{lem:ols}
\begin{align*}
   \hat{\tau} &= \frac{\sum_{s=1}^S \sum_{r=1}^R \sum_{t=1}^D D_{srt}(Y_{srt} - \bar{Y}_{sr} - \bar{Y}_{st} - \bar{Y}_{rt} + \bar{\bar{Y}}_s + \bar{\bar{Y}}_r + \bar{\bar{Y}}_t - \bar{\bar{\bar{Y}}})}{\sum_{s=1}^S \sum_{r=1}^R \sum_{t=1}^T \bigg(\tilde{D}_{srt}\bigg)^2}
\end{align*}

Define $N^{(1)}$ 
\begin{align*}
    & N^{(1)}_{sr} = \sum_{t^{\prime}=1}^T D_{srt^{\prime}} && N^{(1)}_{st} = \sum_{r^{\prime}=1}^R D_{sr^{\prime}t} && N^{(1)}_{rt} = \sum_{s^{\prime}=1}^S D_{s^{\prime}rt}\\
    & N^{(1)}_{s} = \sum_{t^{\prime}=1}^T \sum_{r^{\prime} = 1}^R D_{sr^{\prime}t^{\prime}} && N^{(1)}_{r} = \sum_{t^{\prime}=1}^T \sum_{s^{\prime} = 1}^S D_{s^{\prime}rt^{\prime}} && N^{(1)}_{t} =  \sum_{s^{\prime} = 1}^S \sum_{r^{\prime}=1}^R D_{s^{\prime}r^{\prime}t}\\
    & N^{(1)} = \sum_{s^{\prime}=1}^S \sum_{r^{\prime} = 1}^R \sum_{t^{\prime}=1}^T  D_{s^{\prime}r^{\prime}t^{\prime}} 
\end{align*}

Define $N^{(0)}$ analogously for the control units: $N^{(0)} = \sum_{s^{\prime}=1}^S \sum_{t^{\prime}=1}^T \sum_{r^{\prime} = 1}^R (1 - D_{s^{\prime}r^{\prime}t^{\prime}})$

Expanding the sum and simplifying, we can write the denominator as
\begin{align*}
    \sum_{s=1}^S \sum_{r=1}^R \sum_{t=1}^T \bigg(\tilde{D}_{srt}\bigg)^2 = N^{(1)} &- \frac{1}{T}\sum_{s=1}^S\sum_{r=1}^R (N_{sr}^{(1)})^2  -  \frac{1}{R}\sum_{s=1}^S\sum_{t=1}^T (N_{st}^{(1)})^2 - \frac{1}{S}\sum_{r=1}^R\sum_{t=1}^T (N_{rt}^{(1)})^2\\
    & + \frac{1}{SR} \sum_{t=1}^T (N_{t}^{(1)})^2  + \frac{1}{ST} \sum_{r=1}^R (N_{r}^{(1)})^2  + \frac{1}{RT} \sum_{s=1}^S (N_{s}^{(1)})^2\\
    & -\frac{(N^{(1)})^2}{SRT}
\end{align*}

Re-write the numerator in terms of the double sums
\begin{align*}
    \sum_{r=1}^R \sum_{s=1}^S \sum_{t=1}^T \bigg[ Y_{srt}D_{srt} - \frac{1}{T} \sum_{t^{\prime}=1}^T Y_{srt^{\prime}}D_{srt} - \frac{1}{S} \sum_{s^{\prime}=1}^S Y_{s^{\prime}rt}D_{srt} + \frac{1}{ST} \sum_{s^{\prime}=1}^S \sum_{t^{\prime}=1}^T Y_{s^{\prime}rt^{\prime}}D_{srt}\bigg] - \\ \frac{1}{R}\sum_{r^{\prime}=1}^R \bigg[ Y_{sr^{\prime}t}D_{srt} - \frac{1}{T} \sum_{t^{\prime}=1}^T Y_{sr^{\prime}t^{\prime}}D_{srt} - \frac{1}{S} \sum_{s^{\prime}=1}^S Y_{s^{\prime}r^{\prime}t}D_{srt} + \frac{1}{ST} \sum_{s^{\prime}=1}^S \sum_{t^{\prime}=1}^T Y_{s^{\prime}r^{\prime}t^{\prime}}D_{srt}\bigg]
\end{align*}

Rearranging the sums and cancelling
\begin{align*}
       \frac{1}{SRT}\sum_{r=1}^R \sum_{s=1}^S \sum_{t=1}^T \sum_{s^\prime=1}^S \sum_{t^\prime=1}^T \sum_{r^{\prime} \neq r} D_{srt}\bigg[\left(Y_{srt} - Y_{srt^{\prime}} - Y_{s^{\prime}rt} + Y_{s^{\prime}rt^{\prime}}\right) - \left(Y_{sr^{\prime}t} - Y_{sr^{\prime}t^{\prime}} - Y_{s^{\prime}r^{\prime}t} + Y_{s^{\prime}r^{\prime}t^{\prime}}\right)\bigg]
\end{align*}

Treatment is binary, so $\sum D_{srt} = \sum D_{srt}D_{s^{\prime}rt} + \sum D_{srt}(1-D_{s^{\prime}r t})$

Next, denote the difference-in-difference
\begin{align*}
    \tilde{Y}_{srt}^{(s^{\prime}t^{\prime})} &= Y_{srt} - Y_{s^{\prime}rt} - Y_{srt^{\prime}} + Y_{s^{\prime}rt^{\prime}}
\end{align*}
Note that swapping any one index $s$, $t$ or $r$ (e.g. $s$ for $s^{\prime}$) will flip the sign of the difference-in-differences as well as the triple-difference term $\tilde{Y}_{srt}^{(s^{\prime}t^{\prime})} -  \tilde{Y}_{sr^\prime t}^{(s^{\prime}t^{\prime})}$.

Split the sum using $D_{srt^{\prime}}$, $D_{s^{\prime}rt}$ and $D_{s^{\prime}rt^{\prime}}$
\begin{align*}
       \frac{1}{SRT}\sum_{r=1}^R \sum_{s=1}^S \sum_{t=1}^T \sum_{s^\prime=1}^S \sum_{t^\prime=1}^T \sum_{r^{\prime} \neq r} D_{srt}D_{srt^{\prime}}D_{s^{\prime}rt} D_{s^{\prime}rt^{\prime}}\bigg[ \tilde{Y}_{srt}^{(s^{\prime}t^{\prime})} -  \tilde{Y}_{sr^\prime t}^{(s^{\prime}t^{\prime})}\bigg] + 
       \\ \frac{1}{SRT}\sum_{r=1}^R \sum_{s=1}^S \sum_{t=1}^T \sum_{s^\prime=1}^S \sum_{t^\prime=1}^T \sum_{r^{\prime} \neq r} D_{srt}D_{srt^{\prime}}D_{s^{\prime}rt}(1 - D_{s^{\prime}rt^{\prime}})\bigg[\tilde{Y}_{srt}^{(s^{\prime}t^{\prime})} -  \tilde{Y}_{sr^\prime t}^{(s^{\prime}t^{\prime})}\bigg] + 
       \\ \frac{1}{SRT}\sum_{r=1}^R \sum_{s=1}^S \sum_{t=1}^T \sum_{s^\prime=1}^S \sum_{t^\prime=1}^T \sum_{r^{\prime} \neq r} D_{srt}D_{srt^{\prime}}(1 - D_{s^{\prime}rt})D_{s^{\prime}rt^{\prime}}\bigg[\tilde{Y}_{srt}^{(s^{\prime}t^{\prime})} -  \tilde{Y}_{sr^\prime t}^{(s^{\prime}t^{\prime})}\bigg] + 
       \\ \frac{1}{SRT}\sum_{r=1}^R \sum_{s=1}^S \sum_{t=1}^T \sum_{s^\prime=1}^S \sum_{t^\prime=1}^T \sum_{r^{\prime} \neq r} D_{srt}(1 - D_{srt^{\prime}})D_{s^{\prime}rt} D_{s^{\prime}rt^{\prime}}\bigg[\tilde{Y}_{srt}^{(s^{\prime}t^{\prime})} -  \tilde{Y}_{sr^\prime t}^{(s^{\prime}t^{\prime})}\bigg] + 
       \\ \frac{1}{SRT}\sum_{r=1}^R \sum_{s=1}^S \sum_{t=1}^T \sum_{s^\prime=1}^S \sum_{t^\prime=1}^T \sum_{r^{\prime} \neq r} D_{srt}(1 - D_{srt^{\prime}})D_{s^{\prime}rt}(1 - D_{s^{\prime}rt^{\prime}})\bigg[\tilde{Y}_{srt}^{(s^{\prime}t^{\prime})} -  \tilde{Y}_{sr^\prime t}^{(s^{\prime}t^{\prime})}\bigg] + 
       \\ \frac{1}{SRT}\sum_{r=1}^R \sum_{s=1}^S \sum_{t=1}^T \sum_{s^\prime=1}^S \sum_{t^\prime=1}^T \sum_{r^{\prime} \neq r} D_{srt}D_{srt^{\prime}}(1 - D_{s^{\prime}rt})(1 - D_{s^{\prime}rt^{\prime}})\bigg[\tilde{Y}_{srt}^{(s^{\prime}t^{\prime})} -  \tilde{Y}_{sr^\prime t}^{(s^{\prime}t^{\prime})}\bigg] + 
       \\ \frac{1}{SRT}\sum_{r=1}^R \sum_{s=1}^S \sum_{t=1}^T \sum_{s^\prime=1}^S \sum_{t^\prime=1}^T \sum_{r^{\prime} \neq r} D_{srt}(1 - D_{srt^{\prime}})(1 - D_{s^{\prime}rt}) D_{s^{\prime}rt^{\prime}}\bigg[\tilde{Y}_{srt}^{(s^{\prime}t^{\prime})} -  \tilde{Y}_{sr^\prime t}^{(s^{\prime}t^{\prime})}\bigg] + 
       \\ \frac{1}{SRT}\sum_{r=1}^R \sum_{s=1}^S \sum_{t=1}^T \sum_{s^\prime=1}^S \sum_{t^\prime=1}^T \sum_{r^{\prime} \neq r} D_{srt}(1 - D_{srt^{\prime}})(1 - D_{s^{\prime}rt}) (1 - D_{s^{\prime}rt^{\prime}})\bigg[\tilde{Y}_{srt}^{(s^{\prime}t^{\prime})} -  \tilde{Y}_{sr^\prime t}^{(s^{\prime}t^{\prime})}\bigg] 
\end{align*}

The first term can be shown to be zero as all $Y_{srt}$ and $Y_{sr^{\prime}t}$ will cancel out. Second and fourth terms cancel as we can swap indices $s$ and $s^{\prime}$ to get the other. Fifth and sixth terms also cancel as we can swap indices $t$ and $t^{\prime}$. Under staggered adoption, the seventh term is zero as $D_{srt}(1 - D_{srt^{\prime}})(1 - D_{s^{\prime}rt}) D_{s^{\prime}rt^{\prime}}$ is never equal to $1$. This leaves the third and the eighth terms. The latter corresponds to the ``valid" difference-in-difference (where all comparisons are with control units) while the former is an ``invalid" difference-in-difference as the second difference will involve a future period $t^{\prime} > t$.

\begin{align*}
       \\ \frac{1}{SRT}\sum_{r=1}^R \sum_{s=1}^S \sum_{t=1}^T \sum_{s^\prime=1}^S \sum_{t^\prime=1}^T \sum_{r^{\prime} \neq r} D_{srt}(1 - D_{srt^{\prime}})(1 - D_{s^{\prime}rt}) (1 - D_{s^{\prime}rt^{\prime}})\bigg[\tilde{Y}_{srt}^{(s^{\prime}t^{\prime})} -  \tilde{Y}_{sr^\prime t}^{(s^{\prime}t^{\prime})}\bigg] +
       \\ \frac{1}{SRT}\sum_{r=1}^R \sum_{s=1}^S \sum_{t=1}^T \sum_{s^\prime=1}^S \sum_{t^\prime=1}^T \sum_{r^{\prime} \neq r} D_{srt}D_{srt^{\prime}}(1 - D_{s^{\prime}rt})D_{s^{\prime}rt^{\prime}}\bigg[\tilde{Y}_{srt}^{(s^{\prime}t^{\prime})} -  \tilde{Y}_{sr^\prime t}^{(s^{\prime}t^{\prime})}\bigg] 
\end{align*}

Splitting again on $D_{sr^{\prime}t}$, $D_{s^{\prime}r^{\prime}t}$, $D_{sr^{\prime}t^{\prime}}$ and $D_{s^{\prime}r^{\prime}t^{\prime}}$ and suppressing the six sums for space:

\footnotesize
\begin{align*}
       \\  \bigg[D_{srt}(1 - D_{srt^{\prime}})(1 - D_{s^{\prime}rt}) (1 - D_{s^{\prime}rt^{\prime}})\bigg]\bigg[(1-D_{sr^{\prime}t})(1-D_{s^{\prime}r^{\prime}t})(1-D_{sr^{\prime}t^{\prime}})(1-D_{s^{\prime}r^{\prime}t^{\prime}})\bigg]\bigg[\tilde{Y}_{srt}^{(s^{\prime}t^{\prime})} -  \tilde{Y}_{sr^\prime t}^{(s^{\prime}t^{\prime})}\bigg] +
       \\  \bigg[D_{srt}(1 - D_{srt^{\prime}})(1 - D_{s^{\prime}rt}) (1 - D_{s^{\prime}rt^{\prime}})\bigg]\bigg[(D_{sr^{\prime}t})(1-D_{s^{\prime}r^{\prime}t})(1-D_{sr^{\prime}t^{\prime}})(1-D_{s^{\prime}r^{\prime}t^{\prime}})\bigg]\bigg[\tilde{Y}_{srt}^{(s^{\prime}t^{\prime})} -  \tilde{Y}_{sr^\prime t}^{(s^{\prime}t^{\prime})}\bigg] +
      \\  \bigg[D_{srt}(1 - D_{srt^{\prime}})(1 - D_{s^{\prime}rt}) (1 - D_{s^{\prime}rt^{\prime}})\bigg]\bigg[(1-D_{sr^{\prime}t})(D_{s^{\prime}r^{\prime}t})(1-D_{sr^{\prime}t^{\prime}})(1-D_{s^{\prime}r^{\prime}t^{\prime}})\bigg]\bigg[\tilde{Y}_{srt}^{(s^{\prime}t^{\prime})} -  \tilde{Y}_{sr^\prime t}^{(s^{\prime}t^{\prime})}\bigg] +
     \\  \bigg[D_{srt}(1 - D_{srt^{\prime}})(1 - D_{s^{\prime}rt}) (1 - D_{s^{\prime}rt^{\prime}})\bigg]\bigg[(1-D_{sr^{\prime}t})(1-D_{s^{\prime}r^{\prime}t})(D_{sr^{\prime}t^{\prime}})(1-D_{s^{\prime}r^{\prime}t^{\prime}})\bigg]\bigg[\tilde{Y}_{srt}^{(s^{\prime}t^{\prime})} -  \tilde{Y}_{sr^\prime t}^{(s^{\prime}t^{\prime})}\bigg] +
        \\  \bigg[D_{srt}(1 - D_{srt^{\prime}})(1 - D_{s^{\prime}rt}) (1 - D_{s^{\prime}rt^{\prime}})\bigg]\bigg[(1-D_{sr^{\prime}t})(1-D_{s^{\prime}r^{\prime}t})(1-D_{sr^{\prime}t^{\prime}})(D_{s^{\prime}r^{\prime}t^{\prime}})\bigg]\bigg[\tilde{Y}_{srt}^{(s^{\prime}t^{\prime})} -  \tilde{Y}_{sr^\prime t}^{(s^{\prime}t^{\prime})}\bigg] +
       \\  \bigg[D_{srt}(1 - D_{srt^{\prime}})(1 - D_{s^{\prime}rt}) (1 - D_{s^{\prime}rt^{\prime}})\bigg]\bigg[(D_{sr^{\prime}t})(1-D_{s^{\prime}r^{\prime}t})(D_{sr^{\prime}t^{\prime}})(1-D_{s^{\prime}r^{\prime}t^{\prime}})\bigg]\bigg[\tilde{Y}_{srt}^{(s^{\prime}t^{\prime})} -  \tilde{Y}_{sr^\prime t}^{(s^{\prime}t^{\prime})}\bigg] +
      \\  \bigg[D_{srt}(1 - D_{srt^{\prime}})(1 - D_{s^{\prime}rt}) (1 - D_{s^{\prime}rt^{\prime}})\bigg]\bigg[(D_{sr^{\prime}t})(1-D_{s^{\prime}r^{\prime}t})(D_{sr^{\prime}t^{\prime}})(1-D_{s^{\prime}r^{\prime}t^{\prime}})\bigg]\bigg[\tilde{Y}_{srt}^{(s^{\prime}t^{\prime})} -  \tilde{Y}_{sr^\prime t}^{(s^{\prime}t^{\prime})}\bigg] +
     \\  \bigg[D_{srt}(1 - D_{srt^{\prime}})(1 - D_{s^{\prime}rt}) (1 - D_{s^{\prime}rt^{\prime}})\bigg]\bigg[(D_{sr^{\prime}t})(1-D_{s^{\prime}r^{\prime}t})(1-D_{sr^{\prime}t^{\prime}})(D_{s^{\prime}r^{\prime}t^{\prime}})\bigg]\bigg[\tilde{Y}_{srt}^{(s^{\prime}t^{\prime})} -  \tilde{Y}_{sr^\prime t}^{(s^{\prime}t^{\prime})}\bigg] +
        \\  \bigg[D_{srt}(1 - D_{srt^{\prime}})(1 - D_{s^{\prime}rt}) (1 - D_{s^{\prime}rt^{\prime}})\bigg]\bigg[(1-D_{sr^{\prime}t})(D_{s^{\prime}r^{\prime}t})(D_{sr^{\prime}t^{\prime}})(1-D_{s^{\prime}r^{\prime}t^{\prime}})\bigg]\bigg[\tilde{Y}_{srt}^{(s^{\prime}t^{\prime})} -  \tilde{Y}_{sr^\prime t}^{(s^{\prime}t^{\prime})}\bigg] +
       \\  \bigg[D_{srt}(1 - D_{srt^{\prime}})(1 - D_{s^{\prime}rt}) (1 - D_{s^{\prime}rt^{\prime}})\bigg]\bigg[(1-D_{sr^{\prime}t})(D_{s^{\prime}r^{\prime}t})(1-D_{sr^{\prime}t^{\prime}})(D_{s^{\prime}r^{\prime}t^{\prime}})\bigg]\bigg[\tilde{Y}_{srt}^{(s^{\prime}t^{\prime})} -  \tilde{Y}_{sr^\prime t}^{(s^{\prime}t^{\prime})}\bigg] +       
       \\  \bigg[D_{srt}(1 - D_{srt^{\prime}})(1 - D_{s^{\prime}rt}) (1 - D_{s^{\prime}rt^{\prime}})\bigg]\bigg[(1-D_{sr^{\prime}t})(1-D_{s^{\prime}r^{\prime}t})(D_{sr^{\prime}t^{\prime}})(D_{s^{\prime}r^{\prime}t^{\prime}})\bigg]\bigg[\tilde{Y}_{srt}^{(s^{\prime}t^{\prime})} -  \tilde{Y}_{sr^\prime t}^{(s^{\prime}t^{\prime})}\bigg] +
       \\  \bigg[D_{srt}(1 - D_{srt^{\prime}})(1 - D_{s^{\prime}rt}) (1 - D_{s^{\prime}rt^{\prime}})\bigg]\bigg[(1-D_{sr^{\prime}t})(D_{s^{\prime}r^{\prime}t})(D_{sr^{\prime}t^{\prime}})(D_{s^{\prime}r^{\prime}t^{\prime}})\bigg]\bigg[\tilde{Y}_{srt}^{(s^{\prime}t^{\prime})} -  \tilde{Y}_{sr^\prime t}^{(s^{\prime}t^{\prime})}\bigg] +
           \\  \bigg[D_{srt}(1 - D_{srt^{\prime}})(1 - D_{s^{\prime}rt}) (1 - D_{s^{\prime}rt^{\prime}})\bigg]\bigg[(D_{sr^{\prime}t})(1-D_{s^{\prime}r^{\prime}t})(D_{sr^{\prime}t^{\prime}})(D_{s^{\prime}r^{\prime}t^{\prime}})\bigg]\bigg[\tilde{Y}_{srt}^{(s^{\prime}t^{\prime})} -  \tilde{Y}_{sr^\prime t}^{(s^{\prime}t^{\prime})}\bigg] +
               \\  \bigg[D_{srt}(1 - D_{srt^{\prime}})(1 - D_{s^{\prime}rt}) (1 - D_{s^{\prime}rt^{\prime}})\bigg]\bigg[(D_{sr^{\prime}t})(D_{s^{\prime}r^{\prime}t})(1-D_{sr^{\prime}t^{\prime}})(D_{s^{\prime}r^{\prime}t^{\prime}})\bigg]\bigg[\tilde{Y}_{srt}^{(s^{\prime}t^{\prime})} -  \tilde{Y}_{sr^\prime t}^{(s^{\prime}t^{\prime})}\bigg] +
                   \\  \bigg[D_{srt}(1 - D_{srt^{\prime}})(1 - D_{s^{\prime}rt}) (1 - D_{s^{\prime}rt^{\prime}})\bigg]\bigg[(D_{sr^{\prime}t})(D_{s^{\prime}r^{\prime}t})(D_{sr^{\prime}t^{\prime}})(1-D_{s^{\prime}r^{\prime}t^{\prime}})\bigg]\bigg[\tilde{Y}_{srt}^{(s^{\prime}t^{\prime})} -  \tilde{Y}_{sr^\prime t}^{(s^{\prime}t^{\prime})}\bigg] +
            \\  \bigg[D_{srt}(1 - D_{srt^{\prime}})(1 - D_{s^{\prime}rt}) (1 - D_{s^{\prime}rt^{\prime}})\bigg]\bigg[(D_{sr^{\prime}t})(D_{s^{\prime}r^{\prime}t})(D_{sr^{\prime}t^{\prime}})(D_{s^{\prime}r^{\prime}t^{\prime}})\bigg]\bigg[\tilde{Y}_{srt}^{(s^{\prime}t^{\prime})} -  \tilde{Y}_{sr^\prime t}^{(s^{\prime}t^{\prime})}\bigg] +
\end{align*}
\normalsize

Row 2 cancels with itself as does row 5. Row 4 is ruled out by staggered adoption. Similar argument for rows 12-14. Staggered adoption eliminates rows 8 and 9

\footnotesize
\begin{align*}
       \\  \bigg[D_{srt}(1 - D_{srt^{\prime}})(1 - D_{s^{\prime}rt}) (1 - D_{s^{\prime}rt^{\prime}})\bigg]\bigg[(1-D_{sr^{\prime}t})(1-D_{s^{\prime}r^{\prime}t})(1-D_{sr^{\prime}t^{\prime}})(1-D_{s^{\prime}r^{\prime}t^{\prime}})\bigg]\bigg[\tilde{Y}_{srt}^{(s^{\prime}t^{\prime})} -  \tilde{Y}_{sr^\prime t}^{(s^{\prime}t^{\prime})}\bigg] +
      \\  \bigg[D_{srt}(1 - D_{srt^{\prime}})(1 - D_{s^{\prime}rt}) (1 - D_{s^{\prime}rt^{\prime}})\bigg]\bigg[(1-D_{sr^{\prime}t})(D_{s^{\prime}r^{\prime}t})(1-D_{sr^{\prime}t^{\prime}})(1-D_{s^{\prime}r^{\prime}t^{\prime}})\bigg]\bigg[\tilde{Y}_{srt}^{(s^{\prime}t^{\prime})} -  \tilde{Y}_{sr^\prime t}^{(s^{\prime}t^{\prime})}\bigg] +
       \\  \bigg[D_{srt}(1 - D_{srt^{\prime}})(1 - D_{s^{\prime}rt}) (1 - D_{s^{\prime}rt^{\prime}})\bigg]\bigg[(D_{sr^{\prime}t})(1-D_{s^{\prime}r^{\prime}t})(D_{sr^{\prime}t^{\prime}})(1-D_{s^{\prime}r^{\prime}t^{\prime}})\bigg]\bigg[\tilde{Y}_{srt}^{(s^{\prime}t^{\prime})} -  \tilde{Y}_{sr^\prime t}^{(s^{\prime}t^{\prime})}\bigg] +
      \\  \bigg[D_{srt}(1 - D_{srt^{\prime}})(1 - D_{s^{\prime}rt}) (1 - D_{s^{\prime}rt^{\prime}})\bigg]\bigg[(D_{sr^{\prime}t})(1-D_{s^{\prime}r^{\prime}t})(D_{sr^{\prime}t^{\prime}})(1-D_{s^{\prime}r^{\prime}t^{\prime}})\bigg]\bigg[\tilde{Y}_{srt}^{(s^{\prime}t^{\prime})} -  \tilde{Y}_{sr^\prime t}^{(s^{\prime}t^{\prime})}\bigg] +
       \\  \bigg[D_{srt}(1 - D_{srt^{\prime}})(1 - D_{s^{\prime}rt}) (1 - D_{s^{\prime}rt^{\prime}})\bigg]\bigg[(1-D_{sr^{\prime}t})(D_{s^{\prime}r^{\prime}t})(1-D_{sr^{\prime}t^{\prime}})(D_{s^{\prime}r^{\prime}t^{\prime}})\bigg]\bigg[\tilde{Y}_{srt}^{(s^{\prime}t^{\prime})} -  \tilde{Y}_{sr^\prime t}^{(s^{\prime}t^{\prime})}\bigg] +       
       \\  \bigg[D_{srt}(1 - D_{srt^{\prime}})(1 - D_{s^{\prime}rt}) (1 - D_{s^{\prime}rt^{\prime}})\bigg]\bigg[(1-D_{sr^{\prime}t})(1-D_{s^{\prime}r^{\prime}t})(D_{sr^{\prime}t^{\prime}})(D_{s^{\prime}r^{\prime}t^{\prime}})\bigg]\bigg[\tilde{Y}_{srt}^{(s^{\prime}t^{\prime})} -  \tilde{Y}_{sr^\prime t}^{(s^{\prime}t^{\prime})}\bigg] +
           \\  \bigg[D_{srt}(1 - D_{srt^{\prime}})(1 - D_{s^{\prime}rt}) (1 - D_{s^{\prime}rt^{\prime}})\bigg]\bigg[(D_{sr^{\prime}t})(D_{s^{\prime}r^{\prime}t})(D_{sr^{\prime}t^{\prime}})(1-D_{s^{\prime}r^{\prime}t^{\prime}})\bigg]\bigg[\tilde{Y}_{srt}^{(s^{\prime}t^{\prime})} -  \tilde{Y}_{sr^\prime t}^{(s^{\prime}t^{\prime})}\bigg] +
            \\  \bigg[D_{srt}(1 - D_{srt^{\prime}})(1 - D_{s^{\prime}rt}) (1 - D_{s^{\prime}rt^{\prime}})\bigg]\bigg[(D_{sr^{\prime}t})(D_{s^{\prime}r^{\prime}t})(D_{sr^{\prime}t^{\prime}})(D_{s^{\prime}r^{\prime}t^{\prime}})\bigg]\bigg[\tilde{Y}_{srt}^{(s^{\prime}t^{\prime})} -  \tilde{Y}_{sr^\prime t}^{(s^{\prime}t^{\prime})}\bigg] +
\end{align*}
\normalsize

Applying the same to the terms with $D_{srt}D_{srt^{\prime}}(1 - D_{s^{\prime}rt})D_{s^{\prime}rt^{\prime}}$ and collecting some terms yields an expression for the numerator.
Define $D_{srt}^{(0)} \equiv 1-D_{srt}$.

\begin{align*} 
 & \frac{1}{SRT}\sum_{r=1}^R \sum_{s=1}^S \sum_{t=1}^T \sum_{s^{\prime}=1}^S \sum_{t^{\prime}=1}^T \sum_{r^{\prime} \neq r} \bigg[\tilde{Y}_{srt}^{(s^{\prime}t^{\prime})} -  \tilde{Y}_{sr^\prime t}^{(s^{\prime}t^{\prime})}\bigg] \times \bigg[D_{srt}D_{s^{\prime}rt}^{(0)}D_{srt^{\prime}}^{(0)}D_{s^{\prime}rt^{\prime}}^{(0)} + {D_{srt}D_{s^{\prime}rt}^{(0)}D_{srt^{\prime}}D_{s^{\prime}rt^{\prime}}}\bigg] \times \\ & \bigg[{D_{sr^{\prime}t}^{(0)}D_{s^{\prime}r^{\prime}t}^{(0)}D_{sr^{\prime}t^{\prime}}^{(0)}D_{s^{\prime}r^{\prime}t^{\prime}}^{(0)}} + {D_{sr^{\prime}t}^{(0)}D_{s^{\prime}r^{\prime}t}^{(0)}D_{sr^{\prime}t^{\prime}}D_{s^{\prime}r^{\prime}t^{\prime}}} + {D_{sr^{\prime}t}^{(0)}D_{s^{\prime}r^{\prime}t}D_{sr^{\prime}t^{\prime}}^{(0)}D_{s^{\prime}r^{\prime}t^{\prime}}} + \\ &
{D_{sr^{\prime}t}D_{s^{\prime}r^{\prime}t}D_{sr^{\prime}t^{\prime}}D_{s^{\prime}r^{\prime}t^{\prime}}} + {D_{sr^{\prime}t}D_{s^{\prime}r^{\prime}t}D_{sr^{\prime}t^{\prime}}^{(0)}D_{s^{\prime}r^{\prime}t^{\prime}}^{(0)}} + {D_{sr^{\prime}t}D_{s^{\prime}r^{\prime}t}^{(0)}D_{sr^{\prime}t^{\prime}}D_{s^{\prime}r^{\prime}t^{\prime}}^{(0)}} \bigg] +\\
&  \bigg[ \tilde{Y}_{srt}^{(s^{\prime}t^{\prime})} -  \tilde{Y}_{sr^\prime t}^{(s^{\prime}t^{\prime})}\bigg]\times \bigg[{D_{srt}D_{s^{\prime}rt}^{(0)}D_{srt^{\prime}}^{(0)}D_{s^{\prime}rt^{\prime}}^{(0)}}\bigg] \times \bigg[ {D_{sr^{\prime}t}^{(0)}D_{s^{\prime}r^{\prime}t}D_{sr^{\prime}t^{\prime}}^{(0)}D_{s^{\prime}r^{\prime}t^{\prime}}^{(0)} + D_{sr^{\prime}t}D_{s^{\prime}r^{\prime}t}D_{sr^{\prime}t^{\prime}}D_{s^{\prime}r^{\prime}t^{\prime}}^{(0)}}\bigg] +\\
& \bigg[\tilde{Y}_{srt}^{(s^{\prime}t^{\prime})} -  \tilde{Y}_{sr^\prime t}^{(s^{\prime}t^{\prime})}\bigg]\times \bigg[{D_{srt}D_{s^{\prime}rt}^{(0)}D_{srt^{\prime}}D_{s^{\prime}rt^{\prime}}}\bigg] \times \bigg[ {D_{sr^{\prime}t}^{(0)}D_{s^{\prime}r^{\prime}t}D_{sr^{\prime}t^{\prime}}D_{s^{\prime}r^{\prime}t^{\prime}} + D_{sr^{\prime}t}^{(0)}D_{s^{\prime}r^{\prime}t}^{(0)}D_{sr^{\prime}t^{\prime}}D_{s^{\prime}r^{\prime}t^{\prime}}^{(0)}}\bigg]
\end{align*}

Multiplying numerator and denominator by $SRT$ yields the final expression for $\tau$.

It is helpful to consider an alternate decomposition in which some of the triple-differences terms can be re-written as differences-in-differences as the same difference-in-difference comparison acts as both a ``primary" and a ``placebo"

\footnotesize
\begin{align*}
             \bigg[D_{srt}(1 - D_{srt^{\prime}})(1 - D_{s^{\prime}rt}) (1 - D_{s^{\prime}rt^{\prime}})\bigg]\bigg[(1-D_{sr^{\prime}t})(D_{s^{\prime}r^{\prime}t})(1-D_{sr^{\prime}t^{\prime}})(1-D_{s^{\prime}r^{\prime}t^{\prime}})\bigg]\bigg[(Y_{srt} - Y_{s^{\prime}rt} - Y_{srt^{\prime}} + Y_{s^{\prime}rt^{\prime}})\bigg] \\ + \bigg[D_{srt}(1 - D_{srt^{\prime}})(1 - D_{s^{\prime}rt}) (1 - D_{s^{\prime}rt^{\prime}})\bigg]\bigg[(1-D_{sr^{\prime}t})(D_{s^{\prime}r^{\prime}t})(1-D_{sr^{\prime}t^{\prime}})(1-D_{s^{\prime}r^{\prime}t^{\prime}})\bigg]\bigg[-(Y_{sr^{\prime}t} - Y_{s^{\prime}r^{\prime}t} - Y_{sr^{\prime}t^{\prime}} + Y_{s^{\prime}r^{\prime}t^{\prime}})\bigg]
\end{align*}
\normalsize

Swapping indices $s$ and $r$ in the second expression yields

\footnotesize
\begin{align*}
             \bigg[D_{srt}(1 - D_{srt^{\prime}})(1 - D_{s^{\prime}rt}) (1 - D_{s^{\prime}rt^{\prime}})\bigg]\bigg[(1-D_{sr^{\prime}t})(D_{s^{\prime}r^{\prime}t})(1-D_{sr^{\prime}t^{\prime}})(1-D_{s^{\prime}r^{\prime}t^{\prime}})\bigg]\bigg[(Y_{srt} - Y_{s^{\prime}rt} - Y_{srt^{\prime}} + Y_{s^{\prime}rt^{\prime}})\bigg] \\ +  \bigg[(1-D_{sr^{\prime}t})(D_{s^{\prime}r^{\prime}t})(1-D_{sr^{\prime}t^{\prime}})(1-D_{s^{\prime}r^{\prime}t^{\prime}})\bigg]\bigg[D_{srt}(1 - D_{srt^{\prime}})(1 - D_{s^{\prime}rt}) (1 - D_{s^{\prime}rt^{\prime}})\bigg]\bigg[-(Y_{s^{\prime}rt} - Y_{srt} - Y_{s^{\prime}rt^{\prime}} + Y_{srt^{\prime}})\bigg]\\
             = 2 \bigg[D_{srt}(1 - D_{srt^{\prime}})(1 - D_{s^{\prime}rt}) (1 - D_{s^{\prime}rt^{\prime}})\bigg]\bigg[(1-D_{sr^{\prime}t})(D_{s^{\prime}r^{\prime}t})(1-D_{sr^{\prime}t^{\prime}})(1-D_{s^{\prime}r^{\prime}t^{\prime}})\bigg]\bigg[(Y_{srt} - Y_{s^{\prime}rt} - Y_{srt^{\prime}} + Y_{s^{\prime}rt^{\prime}})\bigg]
\end{align*}
\normalsize

Applying the same to all four of these triple difference terms gives an expression for the numerator in terms of a sum over both 2x2x2 and 2x2 terms.

\begin{align*} 
 & \frac{1}{SRT}\sum_{r=1}^R \sum_{s=1}^S \sum_{t=1}^T \sum_{s^{\prime}=1}^S \sum_{t^{\prime}=1}^T \sum_{r^{\prime} \neq r} \bigg[\text{DiDiD}_{srt}^{(s^\prime r^\prime t^\prime)}\bigg] \times \bigg[D_{srt}D_{s^{\prime}rt}^{(0)}D_{srt^{\prime}}^{(0)}D_{s^{\prime}rt^{\prime}}^{(0)} + {D_{srt}D_{s^{\prime}rt}^{(0)}D_{srt^{\prime}}D_{s^{\prime}rt^{\prime}}}\bigg] \times \\ & \bigg[{D_{sr^{\prime}t}^{(0)}D_{s^{\prime}r^{\prime}t}^{(0)}D_{sr^{\prime}t^{\prime}}^{(0)}D_{s^{\prime}r^{\prime}t^{\prime}}^{(0)}} + {D_{sr^{\prime}t}^{(0)}D_{s^{\prime}r^{\prime}t}^{(0)}D_{sr^{\prime}t^{\prime}}D_{s^{\prime}r^{\prime}t^{\prime}}} + {D_{sr^{\prime}t}^{(0)}D_{s^{\prime}r^{\prime}t}D_{sr^{\prime}t^{\prime}}^{(0)}D_{s^{\prime}r^{\prime}t^{\prime}}} + \\ &
{D_{sr^{\prime}t}D_{s^{\prime}r^{\prime}t}D_{sr^{\prime}t^{\prime}}D_{s^{\prime}r^{\prime}t^{\prime}}} + {D_{sr^{\prime}t}D_{s^{\prime}r^{\prime}t}D_{sr^{\prime}t^{\prime}}^{(0)}D_{s^{\prime}r^{\prime}t^{\prime}}^{(0)}} + {D_{sr^{\prime}t}D_{s^{\prime}r^{\prime}t}^{(0)}D_{sr^{\prime}t^{\prime}}D_{s^{\prime}r^{\prime}t^{\prime}}^{(0)}} \bigg] +\\
& 2 \times \bigg[ \tilde{Y}_{srt}^{(s^{\prime}t^{\prime})} \bigg]\times \bigg[{D_{srt}D_{s^{\prime}rt}^{(0)}D_{srt^{\prime}}^{(0)}D_{s^{\prime}rt^{\prime}}^{(0)}}\bigg] \times \bigg[ {D_{sr^{\prime}t}^{(0)}D_{s^{\prime}r^{\prime}t}D_{sr^{\prime}t^{\prime}}^{(0)}D_{s^{\prime}r^{\prime}t^{\prime}}^{(0)} + D_{sr^{\prime}t}D_{s^{\prime}r^{\prime}t}D_{sr^{\prime}t^{\prime}}D_{s^{\prime}r^{\prime}t^{\prime}}^{(0)}}\bigg] +\\
& 2 \times \bigg[ \tilde{Y}_{srt}^{(s^{\prime}t^{\prime})}\bigg]\times \bigg[{D_{srt}D_{s^{\prime}rt}^{(0)}D_{srt^{\prime}}D_{s^{\prime}rt^{\prime}}}\bigg] \times \bigg[ {D_{sr^{\prime}t}^{(0)}D_{s^{\prime}r^{\prime}t}D_{sr^{\prime}t^{\prime}}D_{s^{\prime}r^{\prime}t^{\prime}} + D_{sr^{\prime}t}^{(0)}D_{s^{\prime}r^{\prime}t}^{(0)}D_{sr^{\prime}t^{\prime}}D_{s^{\prime}r^{\prime}t^{\prime}}^{(0)}}\bigg]
\end{align*}

\section{Two-way clustering} \label{a:two-way}

This section provides alternative versions of the replication results from Section \ref{sec:application} for the effects of the WTO/GATT on trade to address critiques from \textcite{aronow2015cluster} and \textcite{carlson2021dyadic} regarding the improper use of clustered standard errors in studies involving dyads. Most papers in this literature cluster standard errors on the dyad. However, \textcite{aronow2015cluster} notes that this likely underestimates the degree of uncertainty in parameter estimates.

To implement this via a bootstrap approach that is compatible with the imputation estimator recommended in the main text, I use the ``pigeonhole" Bayesian bootstrap outlined in \textcite{owen2012bootstrapping}. For each bootstrap iteration, I sample a weight for each sender and each receiver independently from an $\text{Exponential}(1)$ distribution. Then, I assign a weight to each dyad using the product of the sampled sender and receiver weights, fit a weighted fixed effects regression, and generate the imputed treatment effect estimates. I estimate the standard error using the standard deviation of these bootstrapped estimates and construct 95\% normal confidence intervals.

In general, the results suggest markedly greater uncertainty in effect estimates than acknowledged in the existing literature. Nevertheless, I still find some evidence for failed pre-treatment placebo tests. Moreover, I find considerably large uncertainty in estimates for the effect of WTO/GATT participation, which is likely due to the scarcity of control units on which to fit the imputation model.

\begin{table}[ht!]

\caption{Estimated effects of dyadic WTO membership on log imports - two-way clustered SEs}\label{tab:main_results_twoway}
\centering
\begin{tabular}{l|cc|cc}
& \multicolumn{2}{c|}{Fixed-effects Regression} & \multicolumn{2}{c}{Imputation}\\
& Members & Participants & Members & Participants\\
\hline
Estimate & 	$0.0544$ & $0.0777$ &  $0.169$ & $0.176$\\
Std. Err & $(0.0581)$ & $(0.0737)$ & $(0.132)$ & $(0.169)$ \\
95\% CI & $[-0.0595, 0.168]$ & $[-0.0667, 0.222]$ & $[-0.090, 0.428]$ & $[-0.155, 0.507]$\\

\end{tabular}

\raggedright \footnotesize \textit{Notes:} $371,954$ dyads, $163$ countries, $58$ time-periods. Standard errors are two-way pigeonhole bootstrapped with clustering on sender and receiver. 100 bootstrap iterations.

\end{table}

\begin{figure}[ht!]
    \centering
    \includegraphics[scale=.45]{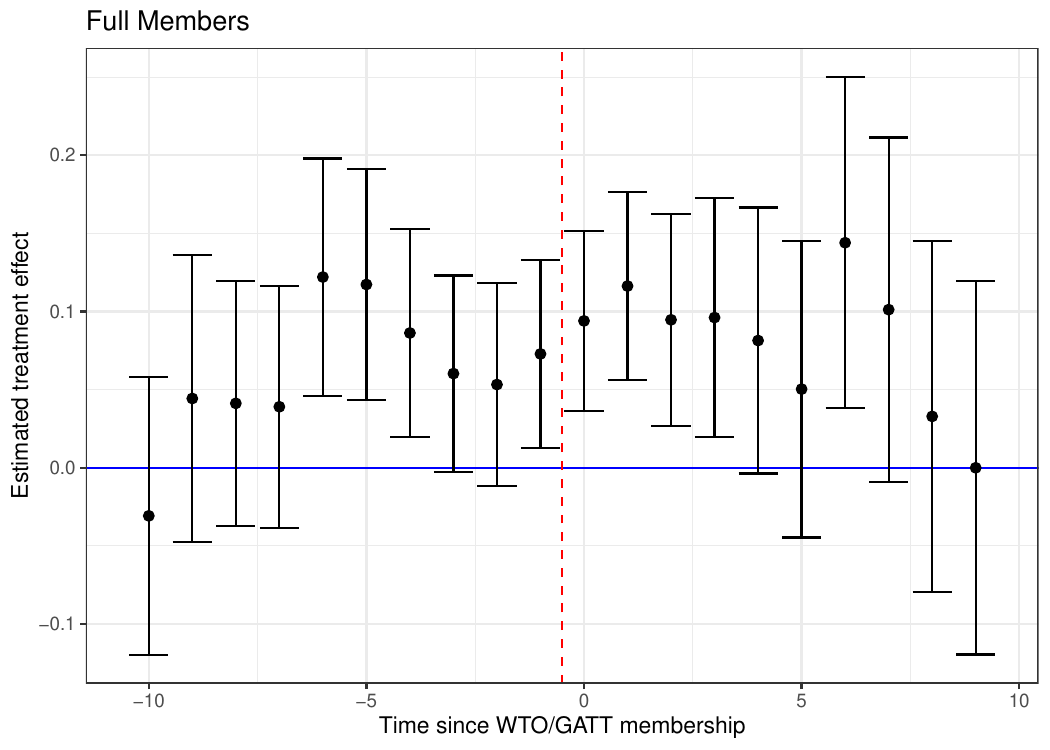}\includegraphics[scale=.45]{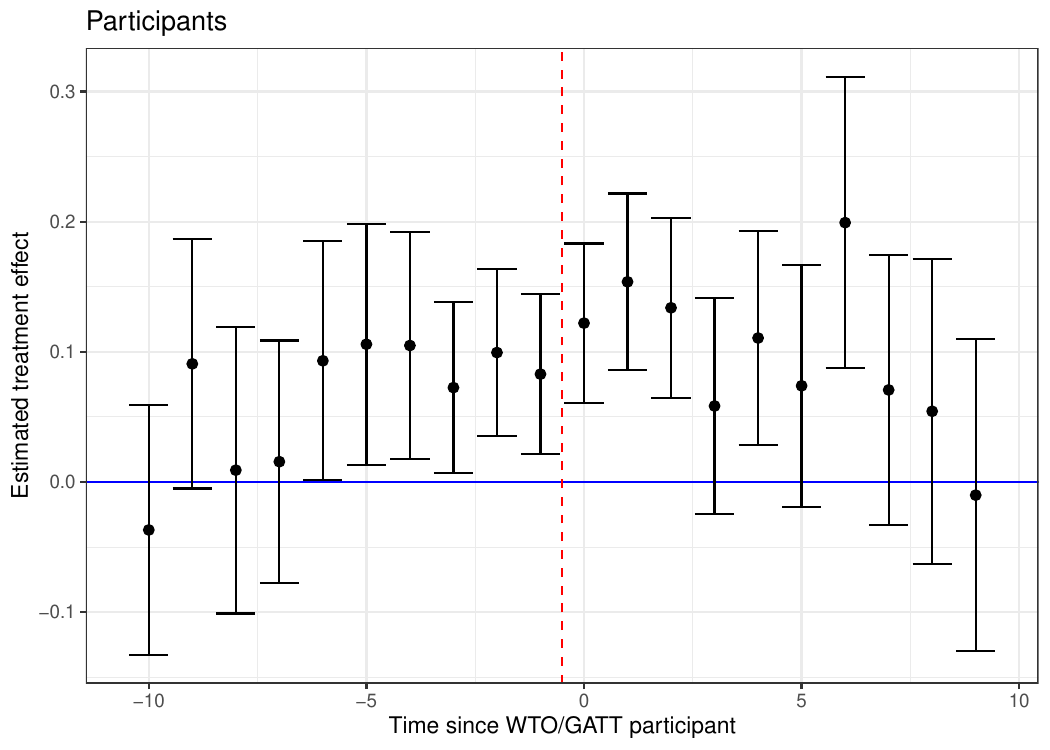}

    \raggedright \footnotesize \textit{Notes:} $371,954$ dyads, $163$ countries, $58$ time-periods. Time = $0$ denotes the first period under treatment. Post-treatment estimates (0 to 10) averaged from imputed effects across all treated units $t$ periods after treatment adoption. Pre-treatment estimates (-10 to -1) are held-out placebo tests. Bars denote 95\% two-way cluster pigeonhole-bootstrapped CIs clustered on sender and receiver. 100 bootstrap iterations.
    
    \caption{Estimated effects of dyadic WTO membership on log imports - effects over time up to 10 years post-treatment - Two-way clustered SEs}
    \label{fig:eventplot_twoway}
\end{figure}

\begin{figure}[ht!]
    \centering
    \includegraphics[scale=.45]{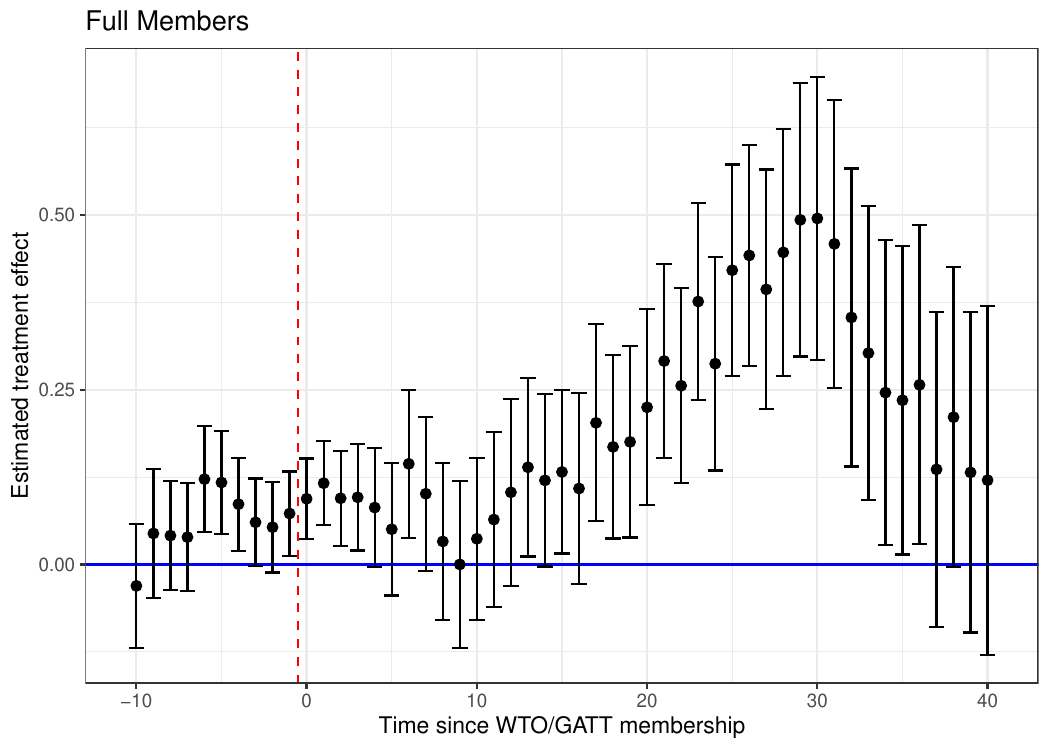} \includegraphics[scale=.45]{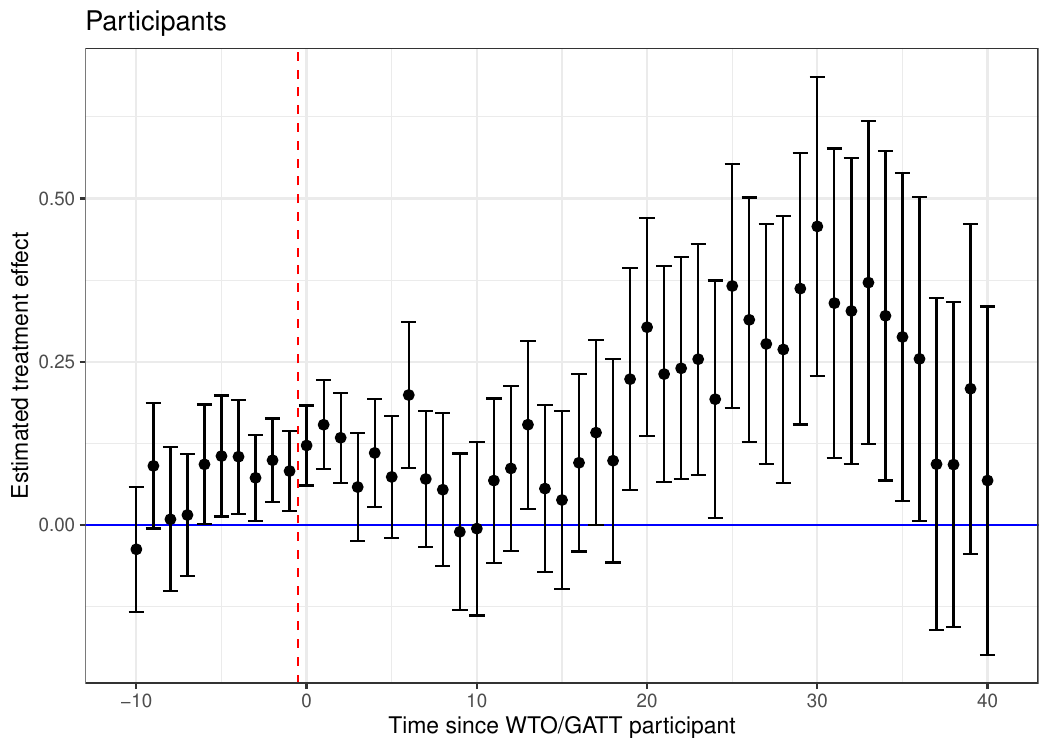}
    
    \raggedright \footnotesize \textit{Notes:} $371,954$ dyads, $163$ countries, $58$ time-periods. Time = $0$ denotes the first period under treatment. Post-treatment estimates (0 to 40) averaged from imputed effects across all treated units $t$ periods after treatment adoption. Pre-treatment estimates (-10 to -1) are held-out placebo tests. Bars denote 95\% two-way cluster pigeonhole-bootstrapped CIs clustered on sender and receiver. 100 bootstrap iterations.
    
    \caption{Estimated effects of dyadic WTO membership on log imports - effects over time up to 40 years post-treatment  - Two-way clustered SEs}
    \label{fig:eventplot_long_twoway}
\end{figure}

\end{appendices}
\end{document}
%%%%%%%%%%%%%%%%%%%%%%%%%%%%%%%%%%%%%%%%%%%%%%%%%%%%%%%%%%%%%%%%%%%%%%%%%%%%%%%%
% == document ends here
%%%%%%%%%%%%%%%%%%%%%%%%%%%%%%%%%%%%%%%%%%%%%%%%%%%%%%%%%%%%%%%%%%%%%%%%%%%%%%%%

DD